\documentclass[onecolumn,astrosymb]{aastex631}

\usepackage{multirow}
\usepackage{amssymb}
\usepackage{xcolor}
\usepackage{hyperref}

\shorttitle{Discovery of Brown Dwarf HIP 17453 B}
\shortauthors{Brinjikji et al.}

\begin{document}

\title{The Companions to B and A Stars Snapshot (C-BASS) Survey: I. Discovery of a Young Brown Dwarf Companion to HIP 17453}

\correspondingauthor{Marah Brinjikji}
\email{mbrinjik@nd.edu}

\author[0000-0002-0457-2941]{Marah Brinjikji}
\affiliation{Department of Physics and Astronomy, University of Notre Dame, 225 Nieuwland Science Hall, Notre Dame, IN 46556, USA}
\affiliation{School of Earth and Space Exploration, Arizona State University, 781 Terrace Mall, Tempe, AZ 85287, USA}

\author[0000-0002-9156-9651]{Adam J. R. W. Smith}
\affiliation{Department of Astronomy, New Mexico State University, 1320 Frenger Mall, Las Cruces, NM 88003, USA}

\author[0000-0001-9004-803X]{Jenny Patience}
\affiliation{School of Earth and Space Exploration, Arizona State University, 781 Terrace Mall, Tempe, AZ 85287, USA}

\author[0000-0001-6975-9056]{Eric L. Nielsen}
\affiliation{Department of Astronomy, New Mexico State University, 1320 Frenger Mall, Las Cruces, NM 88003, USA}

\author[0000-0003-2828-0334]{Austin Ware}
\affiliation{School of Earth and Space Exploration, Arizona State University, 781 Terrace Mall, Tempe, AZ 85287, USA}

\author[0000-0002-9142-6378]{Jorge A. Sanchez}
\affiliation{School of Earth and Space Exploration, Arizona State University, 781 Terrace Mall, Tempe, AZ 85287, USA}

\author[0000-0002-4918-0247]{Robert De Rosa}
\affiliation{European Southern Observatory, Alonso de Cordova 3107, Vitacura, Casilla 19001, Santiago, Chile}

\author[0009-0006-7042-4599]{Jasmine Garani}
\affiliation{Department of Astronomy and Planetary Science, Northern Arizona University, 1440 S. Knoles Dr., Flagstaff, AZ 86011, USA}

\author[0000-0002-1420-1837]{Emma Softich}
\affiliation{Department of Astronomy \& Astrophysics, University of California San Diego, 9500 Gilman Drive, La Jolla, CA 92093, USA}

\author[0000-0003-2232-7664]{Michael C. Liu}
\affiliation{Institute for Astronomy at the University of Hawai`i at M\={a}noa, 2680 Woodlawn Drive, Honolulu, HI 96822, USA}

\author[0000-0003-1212-7538]{Bruce A. Macintosh}
\affiliation{Department of Astronomy and Astrophysics, University of California, Santa Cruz, 1156 High Street, Santa Cruz, CA 95064, USA}

\author[0000-0002-0792-3719]{Thomas M. Esposito}
\affiliation{SETI Institute, Carl Sagan Center, 339 Bernando Ave Suite 200, Mountain View, CA 94043, USA}
\affiliation{Department of Astronomy, University of California Berkeley, 501 Campbell Hall $\#$3411, Berkeley, CA 94720, USA}

\author[0000-0003-2649-2288]{Brendan P. Bowler}
\affiliation{Department of Physics, University of California Santa Barbara, Broida Hall, Santa Barbara, CA 93106, USA}

\author[0000-0003-0562-1511]{William M. J. Best}
\affiliation{Department of Astronomy, University of Texas at Austin, 2511 Speedway, Austin, TX 78705, USA}

\author[0000-0002-3199-2888]{Sarah Blunt}
\affiliation{Department of Astronomy and Astrophysics, University of California, Santa Cruz, 1156 High Street, Santa Cruz, CA 95064, USA}

\author[0000-0001-9823-1445]{Trent J. Dupuy}
\affiliation{Institute for Astronomy, University of Edinburgh, Royal Observatory, Edinburgh, EH9 3HJ, UK}

\author[0000-0001-9994-2142]{Justin Hom}
\affiliation{Steward Observatory and Department of Astronomy, University of Arizona, 933 N Cherry Avenue, Tucson AZ 85721}

\author[0000-0003-3906-9518]{Jessica Klusmeyer}
\affiliation{Department of Astronomy, New Mexico State University, 1320 Frenger Mall, Las Cruces, NM 88003, USA}

\author[0000-0001-7016-7277]{Franck Marchis}
\affiliation{SETI Institute, Carl Sagan Center, 339 Bernando Ave Suite 200, Mountain View, CA 94043, USA}

\author[0000-0002-9242-9052]{Jayke S. Nguyen}
\affiliation{Department of Astronomy \& Astrophysics, University of California San Diego, 9500 Gilman Drive, La Jolla, CA 92093, USA}

\author[0000-0003-2461-6881]{Anne E. Peck}
\affiliation{Department of Astronomy, New Mexico State University, 1320 Frenger Mall, Las Cruces, NM 88003, USA}

\author[0000-0001-6041-7092]{Mark W. Phillips}
\affiliation{Institute for Astronomy, University of Edinburgh, Royal Observatory, Edinburgh, EH9 3HJ, UK}

\author[0009-0008-9687-1877]{William Roberson}
\affiliation{Department of Astronomy, New Mexico State University, 1320 Frenger Mall, Las Cruces, NM 88003, USA}

\author[0000-0003-2233-4821]{Jean-Baptiste Ruffio}
\affiliation{Department of Astronomy \& Astrophysics, University of California San Diego, 9500 Gilman Drive, La Jolla, CA 92093, USA}

\author[0000-0001-5684-4593]{William Thompson}
\affiliation{Herzberg Institute of Astrophysics, National Research Council of Canada, 5071 West Saanich Rd, Victoria, BC V9E 2E7, Canada}

\author[0000-0001-8269-324X]{Zahed Wahhaj}
\affiliation{European Southern Observatory, Alonso de Cordova 3107, Vitacura, Casilla 19001, Santiago, Chile}

\author[0000-0001-7062-815X]{Samuel Walker}
\affiliation{Institute for Astronomy at the University of Hawai`i at M\={a}noa, 2680 Woodlawn Drive, Honolulu, HI 96822, USA}

\author[0000-0003-0774-6502]{Jason J. Wang}
\affiliation{Department of Physics and Astronomy, Northwestern University, 2145 Sheridan Road, Evanston, IL 60208-3112}
\affiliation{Center for Interdisciplinary Exploration and Research in Astrophysics, Northwestern University,  1800 Sherman Ave, Evanston, IL 60201}

\author[0000-0003-1705-5991]{Patrick A. Young}
\affiliation{School of Earth and Space Exploration, Arizona State University, 781 Terrace Mall, Tempe, AZ 85287, USA}

\begin{abstract}
\noindent We report the detection of a new brown dwarf companion to HIP 17453 A, a chemically peculiar A0V star located at a distance of 81 pc. HIP 17453 A was observed with high-resolution adaptive optics imaging using the Near-Infrared Camera 2  on the Keck II telescope as part of the Companions to B and A Stars Snapshot (C-BASS) survey over a ten-year baseline, revealing the presence of a companion with proper motion consistent with the primary. We estimate the age of the HIP 17453 system as 280 $\pm$ 125 Myr, and with follow-up intermediate resolution (R$\sim$1800) spectroscopic observations with the Gemini Near Infra-Red Spectrograph (GNIRS) on the Gemini-North telescope, we found the spectrum of HIP 17453 B to be consistent with a spectral type of L2 $\pm$ 1. Through interpolation of Sonora Diamondback evolutionary models, we calculate an effective temperature of $1953^{+84}_{-78}$ K and mass of $53^{+10}_{-8}$ $M_{Jup}$ for HIP 17453 B, which corresponds to a mass ratio of $q = 0.024 \pm 0.004$ for the HIP 17453 system. With its intermediate mass and young age, HIP 17453 B joins a small set of benchmark brown dwarf companions {around early-type stars} that are suitable for follow-up atmospheric and evolutionary studies.
\end{abstract}


\section{Introduction} \label{sec:intro}

Brown dwarfs occupy a critical position in the continuum between the lowest-mass stars and the most massive exoplanets, providing a natural connection for studying atmospheric and structural properties across the stellar–substellar boundary. {The boundary between planets and brown dwarfs has conventionally been drawn at the deuterium-burning minimum mass ($\sim$13 $M_{\rm Jup}$), defining any object above that mass as a brown dwarf regardless of where it formed and any lower-mass object orbiting a star as a planet \citep[e.g.][]{basri2000a,burr01,kirk05}. However, this definition is physically arbitrary, as deuterium burning has no impact on the evolution of an object, in contrast to hydrogen burning, which distinguishes between brown dwarfs and stars \citep{chabrier14}. What further complicates this mass-based definition is that the exact mass at which deuterium ignites is not fixed at 13 $M_{\rm Jup}$ but instead depends on other factors including helium abundance, initial deuterium content, and metallicity \citep{spieg11}.}

{An alternative distinction between planets and brown dwarfs defines these objects based on how they formed rather than by mass alone. Planets are theorized to form within a circumstellar disk primarily through processes including core accretion, where dusts and solids in the disk coagulate into a core, which above a critical mass can accrete a gaseous envelope, and gravitational instability, where the disk fragments and collapses \citep{dr23,krat10}. By contrast, brown dwarfs are believed to form more like low-mass stars through the gravitational collapse and fragmentation of molecular cloud cores \citep[e.g.][]{w07,w18}. These two distinctions between planets and brown dwarfs do not map cleanly onto each other, as objects forming via core accretion or gravitational instability in disk can theoretically grow beyond the deuterium-burning mass limit, making them brown dwarfs by the mass definition despite forming like planets \citep{mol12,bod13}, while cloud core fragmentation can theoretically produce companions with masses below the deuterium-burning limit \citep{whit06}. This mismatch motivates treating formation mechanism rather than mass alone as the defining dividing line between the brown dwarf and planet populations.}

Radial velocity companion demographic surveys around Sun-like stars have revealed the “brown dwarf desert,” in which {the frequency of brown dwarf companions} in close orbits is much lower than both the planetary companion and stellar companion frequency \citep{gl06}. While surveys have uncovered more brown dwarf companions at wider orbital distances (e.g. \citealt{nielsen12,now17,GPIES,gri21,vow25}), the brown dwarf desert persists at close separations \citep{troup16}. This suggests that different formation mechanisms may dominate at different separations from the host star, leading to questions about the origins of brown dwarfs and how they may differ from their planetary and stellar counterparts.

Theoretical models for the properties and evolution of substellar objects have been extensively developed and applied to both exoplanet and brown dwarf atmospheres \citep{DUSTY00,baraffe03,btsettl10,drift,sm08,sonoradiamondback}. 
These models incorporate various combinations of physical and chemical processes, including condensate cloud formation, disequilibrium chemistry, and vertical mixing, and serve as the boundary conditions for evolutionary models. {These evolutionary cooling tracks model how the properties of the objects change over time, helping predict the mass–luminosity–age relations for substellar objects} \citep{BTSettl,bhac,sonoradiamondback}. 
{Brown dwarfs can serve as higher signal-to-noise benchmarks for exoplanets, }as they are often uncontaminated by host star light and have signal-to-noise ratios higher than those for directly imaged exoplanets of similar temperatures and ages. 
In particular, isolated and wide-orbit brown dwarfs avoid the effects of strong stellar irradiation \citep{fortney2008,show2008,amaral25}, tidal locking and permanent hot spots \citep{sg02,p19}, and atmospheric escape processes, all of which may significantly modify the upper atmospheres of close-in planets \citep{lec04,yelle04,murr09,sing19}. 

Observing brown dwarfs as companions to main-sequence stars offers a number of advantages over observing isolated brown dwarfs. {The ages of main-sequence host stars can often be estimated (e.g. from stellar activity, rotation, kinematics, or comparing to stellar evolution tracks), allowing the luminosities of their brown dwarf companions to be placed on mass–luminosity–age evolutionary tracks. An age estimate breaks the degeneracy between mass and age inherent to substellar evolution, in which objects of markedly different masses can exhibit similar luminosities at different ages, and provides an estimate of the mass for the brown dwarf companion} \citep{burr01}. 
{The demographics of these systems provide empirical tests for competing formation models, distinguishing between planetary-like formation within a protoplanetary disk and binary-like formation via fragmentation of the original molecular cloud that birthed the primary star. While core accretion is considered the primary pathway for giant planets at small separations, brown dwarf companions are thought to originate either through disk instability (which would require massive disks to reach substellar masses) or turbulent fragmentation in a scaled-down version of star formation} \citep{w07,krat10,dr23}. 

Searches for these companions can be broadly divided into unbiased surveys and targeted programs that leverage prior knowledge of host star accelerations from astrometry or radial velocity data \citep{currie21,franson23,rob23}. While targeted searches can yield high efficiency in specific parameter spaces, such as high-mass or close-in systems with detectable dynamical signatures, they are limited in geometric sensitivity and cannot provide unbiased statistical constraints on the overall brown dwarf companion population.

In this work, we present the first discovery of a substellar companion from the Companions to B and A Stars Snapshot (C-BASS) Survey, a program designed to characterize the demographics of wide-orbit companions to early-type stars, {with spectral types ranging from B3–A9}. We outline the survey design, including selection criteria and observational methodology, before describing the properties of the newly identified brown dwarf companion to the A0 star HIP 17453 A. Section \ref{sec:cbass} describes the C-BASS survey, while Section \ref{sec:host} details the properties of HIP 17453 A. Sections \ref{sec:hcobs}–\ref{sec:spec} describe the high-contrast imaging and spectroscopy observations, data reduction, and results. In Section \ref{sec:disc}, we discuss the properties of the new companion, place it in context with other known companions, and consider the implications for occurrence rates to early-type stars. 

\section{The C-BASS Survey} \label{sec:cbass}

The Companions to B and A Stars Snapshot (C-BASS) Survey is a high-contrast imaging survey aimed at increasing the number of brown dwarf companions known around intermediate-mass stars ($\gtrsim$1.5 M$_{\sun}$). The goal of this survey is to set constraints on the overlap between planet formation and binary star formation---and to determine how brown dwarfs fit into both paradigms---from changes in occurrence rate as a function of companion mass. This survey complements direct imaging demographic surveys of exoplanets such as GPIES \citep{GPIES} or SHINE \citep{vigan21} by expanding the number of stars and probing wider separations. Increasing the number of substellar companions to intermediate-mass stars makes it possible to examine key demographic trends: how companion occurrence rate varies as a function of stellar host mass, orbital semi-major axis, and companion mass. In particular, the trends across companion mass from stellar binaries, to brown dwarf companions, to planetary-mass companions can indicate differences in formation mechanisms (e.g. \citealt{bowler20,GPIES}).

The observing strategy of the C-BASS Survey focused on observational efficiency to maximize the number of B and A stars imaged. As these intermediate-mass stars have shorter main-sequence lifetimes ($\lesssim$2 Gyr) than Solar-type stars, their substellar companions will generally be younger and so brighter and easier to detect. Also, while the deep contrasts at small angular separations reached by direct imaging surveys for giant exoplanets ($\sim$15 magnitudes at $\sim$0.5$^{\prime\prime}$, \citealt{macintosh15}) require $\sim$1 hour of observing time, similar contrasts at wider separations are possible with much shorter exposure times. For example, five minutes of on-sky $K_S$ imaging with Keck/NIRC2 reaches 5$\sigma$ contrasts of 11 magnitudes at 1$^{\prime\prime}$ and 13 magnitudes at 3$^{\prime\prime}$. Thus, these shorter observations are sensitive mainly to brown dwarf companions beyond $\sim$1$^{\prime\prime}$ around $\sim$500 Myr stars, compared to a one-hour observation which would be sensitive to planetary-mass companions within 1" around $\lesssim$100 Myr moving group stars. These snapshot observations do not benefit from Angular Differential Imaging \citep[ADI;][]{marois2006,liu10} due to the limited field rotation during the short exposure time, though Reference Differential Imaging \citep[RDI;][]{lafreniere2009,soummer2011,gm2016} is more feasible given the larger number of stars observed in a single night.

The C-BASS target list was chosen from $Hipparcos$ stars within 200 pc, parallax error less than 5\%, color of $-0.2 \le B-V \le 0.4$, and spectral type of B or A. Wide binaries were removed from the target list, as were stars that had significant high contrast imaging observations in the archives (e.g. Keck/NIRC2 or VLT/NACO). Younger stars, as determined from the color-magnitude diagram position using the method described in \citet{nici2}, were prioritized for observations. Most first epoch observations of C-BASS observations were carried out over four nights with Keck/NIRC2 in 2013 and 2014, with 211 stars observed in these four runs. Second epoch observations took place mainly in 2023 and 2024; this decade-long baseline ensured that common proper motion companions could be differentiated from background stars at high confidence \citep{nielsen17}.

In this paper we describe the first substellar object detected by the C-BASS Survey, HIP 17453 B. In future papers we will present stellar companions detected around these stars (Smith et al., in prep) and demographics results from the full sample. 

\section{Host Star Properties}\label{sec:host}

\begin{deluxetable}{cc}
\tablecaption{Properties of HIP 17453 A \label{tab:objects}}
\tablenum{1}
\tablehead{
\colhead{Property} & \colhead{Value}
}
\startdata
Gaia DR3$^a$                      & 63478624499091968 \\
R.A. (ICRS)$^a$                   & 03:44:28.20 \\
Dec. (ICRS)$^a$                   & +20:55:43.45  \\
Spectral Type$^b$                 & A0Vp($\lambda$ Boo)  \\
Distance (pc)$^c$                 & $80.50^{+0.35}_{-0.29}$ \\
$J$ (mag)$^d$                     & 5.995 $\pm$ 0.018 \\
$H$ (mag)$^d$                     & 6.053 $\pm$ 0.023 \\
$K_s$ (mag)$^d$                   & 6.031 $\pm$ 0.023 \\
Median Age (Myr)$^e$              & 280 $\pm$ 125 \\
Mass ($M_{\odot}$)$^e$            & 2.08 $\pm$ 0.09 \\
T$_{eff}$ (K)$^e$                 & 9400 $\pm$ 250 \\
{[}M/H{]}$^e$                     & -0.06 $\pm$ 0.1 \\
$\textit{v} sin (\textit{i})$$^e$ & 123.76 $\pm$ 12
\enddata
\tablerefs{
$a$: \citet{gaiadr3} $b$: \citet{am1995} $c$: \citet{bailer-jones2021} $d$: \citet{2mass} $e$: \textit{This work}
}
\end{deluxetable}

Originally observed as part of the C-BASS survey, HIP 17453 A is a chemically peculiar $\lambda$ Boo A0V star \citep{am1995} {located at a distance of $80.50^{+0.35}_{-0.29}$ pc \citep{bailer-jones2021}.} The $\lambda$ Boo class of objects was first identified by \citet{morgan43} and is {characterized by under-abundant iron peak metals in the spectrum of the star, while volatile elements like C, N, O and S retain solar abundances} \citep{murphy2015}. HIP 17453 A was first identified as a potential chemically peculiar $\lambda$ Boo-type star by \citet{am1995} during their study on the v \textit{sin} i of A-type stars \citep{am1995}. Later studies debated the classification of HIP 17453 A as a $\lambda$ Boo star, citing only moderate under-abundance of metals \citep{andr2002}. \citet{murphy2015} recommend that HIP 17453 A be considered a member of the $\lambda$ Boo class of stars based on a new spectral analysis of the object. From the PASTEL catalogue of stellar atmospheric parameters, HIP 17453 A has T$_{eff}$ = 9500 K, $log$~$g$ (cgs) = 4.3, and a subsolar metallicity of [Fe/H] = -0.67\,dex \citep{pastel,andr2002,zorec2012}. HIP 17453 A had not been previously observed by speckle imaging or AO searches for comoving substellar and stellar companions \citep[e.g.,]{hart84,shaya11,VAST3}, and HIP 17453 A is listed as single by the Washington Double Star Catalog \citep{WDS}. One substellar companion was identified in the AO survey presented in this study, as shown in Figure \ref{fig:17453obs}. A study of the infrared excesses of \textit{Hipparcos} stars revealed no excess infrared emission in the spectral energy distribution of HIP 17453 A \citep{mcdonald2012}. This star is also not known to be a part of any spectroscopic binary system \citep{sb9}. 

We also tested for the presence of an undetected spectroscopic binary using the Apache Point Observatory Astrophysical Research Consortium Echelle Spectrograph (ARCES). The wavelength region around 5200 \AA{} includes a cluster of lines which are strong in A0 stars, including Mg I at 5167 \AA{}, Fe II at 6169 \AA{}, Mg I at 5173 \AA{}, and at Mg I 5183 \AA{}, as well as several weaker lines \citep{and67,kur75,bar2000,ald07,raas98,bark05,K13} which produce a compound feature about 10 \AA{} wide and 5-8\% deep when accounting for the 120 km/s \textit{v} sin (\textit{i}) of HIP 17453 A. This same region is host to many strong ($>$10\%) absorption features in cooler G-type stars. Accounting for magnitude differences, these features, if present, should be detectable at the $\sim$3-5\% level in the combined spectrum, and would appear distinct from the Mg I/Fe II compound feature of the A0 spectrum. We performed the test by creating a compound model spectrum through a weighted sum of two model spectra from \citet{co14}, with the weights of each components controlled by their theoretical relative V magnitudes \citep{pm13}, at a range of relative radial velocity (RV) from -500 $<$ $\Delta$RV $<$ 500 km/s. We tested two hypotheses: an undetected A0 star, and an undetected G0 star. We were unable to find any evidence to support either hypothesis in our data, leading to the conclusion that HIP 17453 A is not in a binary stellar system.

Determining the physical properties of the host stars in the C-BASS sample is critical in understanding the empirical characteristics of any discovered companions, as well as in determining the overall sensitivity of the survey. The ages for B and A stars generally cannot be determined through the typical methods of gyrochronology or chromospheric activity studies, and instead are most often determined through comparison of the physical properties of the star to stellar evolution models. Literature age calculations for HIP 17453 A resulted in ages of 356 Myr, as determined from a study of how the ages of O–F spectral type stars vary depending on kinematics, calculated from the position on a Hertzprung-Russell diagram relative to stellar isochrones \citep{gontcharov2012}, and 385 $\pm$ 40 Myr, as determined from interpolation of photometry onto model grids and comparison with theoretical stellar isochrones \citep{david2015}. 
The procedure originally set in \citet{nici2} and updated in \citet{GPIES} was utilized to determine the age and mass posteriors for the full C-BASS sample using position on the color-magnitude diagram (CMD), as A-stars evolve on the CMD over a period of $<$600 Myr \citep{brandt15}. The \citet{nici2} method utilizes Bayesian inference to estimate age and mass likelihoods from isochrones on the CMD using the age, mass and metallicity of the star as priors (as listed in Table \ref{tab:objects}); \citet{GPIES} updated the Bayesian posteriors to include initial rotation, inclination of the rotation axis, and parallax, and the priors to include rotation rate, inclination and parallax. This study uses the \citet{GPIES} method of combining the MESA stellar evolutionary models, ATLAS9 atmosphere models, and \textit{Gaia} photometry (using \textit{Tycho2} for stars too bright for \textit{Gaia}) to determine the ages of the stars in the C-BASS sample. For the full explanation on this Bayesian method of determining ages, see Section 3.2 from \citet{nici2} and Section 2.1.1 from \citet{GPIES}. The age and mass of HIP 17453 A are estimated as 280 $\pm$ 125 Myr and 2.08 $\pm$ 0.09 $M_{\odot}$ based on the position of the star on the CMD. The full properties of HIP 17453 A are detailed in Table \ref{tab:objects}. 

\section{High-Contrast Imaging Observations, Data Reduction, and Results} \label{sec:hcobs}
\subsection{Imaging Observations} \label{subsec:nirc2obs}
HIP 17453 A was observed as part of the full C-BASS survey by the Keck II telescope at the W. M. Keck Observatory. For all targets in the survey, the Near-Infrared Camera 2 (NIRC2) was used in vertical angle mode with the narrow-field camera and the instrument field rotator turned off alongside the AO system in natural guide star (NGS) mode \citep{NGS}. The initial companion search observations for HIP 17453 A were obtained on 18 Dec 2013 with an instrument setup designed to attenuate the bright star with a 600 milliarcsecond partially-transparent coronagraphic spot with 0.22\% transparency in \textit{K}–band \citep{bowler2015} in order to achieve the contrast necessary to image substellar companions. The narrow-field camera with a pixel scale of 9.952 $\pm$ 0.002 mas/pixel \citep{yelda2010} appropriate for diffraction-limited imaging was selected, along with a \textit{$K_s$}–band filter which covers the wavelength range of enhanced AO performance and is near the peak of emission for brown dwarfs. The first epoch imaging sequence consisted of 3 short exposures of 3 coadds of 10s each and a set of 5 longer 60s exposures with a single coadd. A visual inspection of these observations of HIP 17453 A revealed a candidate companion at an angular separation of $\sim$3$^{\prime\prime}$ as shown in Figure \ref{fig:17453obs}. Details of the observations are given in Table \ref{tab:obs}. {The contrast curve from the first epoch observations is displayed in Figure \ref{fig:contrast_curve}, showing the sensitivity of the C-BASS survey to wide-separation brown dwarf companions.}

A second epoch of  \textit{$K_s$}–band was obtained on 18 Feb 2024 to test for common proper motion for a candidate companion identified in the first epoch of imaging. For data taken after April 2015, the pixel scale of the detector is 9.971 $\pm$ 0.004 mas/pixel using the narrow-field camera \citep{service2016}. The sequence of images obtained in the second epoch matched the approach of the first epoch, combining coronagraphic short exposures and longer exposures. After astrometric analysis confirmed the co-moving nature of the detected companion (see Fig. \ref{fig:CPM}), additional epochs of imaging were obtained during one night in Oct 2024 and another night in Oct 2025 to determine more precise photometry at $K_s$- and $J$-band. For the third and fourth epoch observations of HIP 17453 A, unsaturated images of the star were taken without a coronagraphic mask by using smaller detector subarrays which enabled very short 0.011s exposure times. The observation sequence was altered during the Oct 2024 and 2025 nights; unsaturated exposures of the host star are required for accurate photometry and astrometry of any detected companions as the unsaturated images of the star through the coronagraph returned variable results for the star magnitude. In addition to the short subarray images, 60 second exposures (10 times longer than the unsaturated exposures) using the full detector were then taken to obtain detections of fainter companions with higher signal-to-noise. Details of the Oct 2024 and 2025 follow-up observations of HIP 17453 A are also summarized in Table \ref{tab:obs}.

\begin{deluxetable}{ccccccccc}
\tablecaption{Keck/NIRC2 Observations of HIP 17453 A \label{tab:obs}}
\tablenum{2}
\tablehead{
    \colhead{\multirow{2}{*}{Epoch}} & \colhead{{Date}} & \colhead{\multirow{2}{*}{Band}} & \colhead{Exposure Time} & \colhead{Number of} & \colhead{Number of} & \colhead{\multirow{2}{*}{Coronagraph}} & \colhead{Array} & \colhead{Observation} \\ \\[-0.7cm]
      \colhead{} & \colhead{(UTC)} & \colhead{} & \colhead{(s)} & \colhead{Coadds} & \colhead{Exposures} & \colhead{} & \colhead{Size} & \colhead{Type}
}
\startdata
\multirow{2}{*}{1} & \multirow{2}{*}{2013 Dec 18} & \multirow{2}{*}{$K_s$} & 10 & 3 & 3 & 600mas & 1024 & Unsaturated \\
{} & {} & {} & 60 & 1 & 5 & 600mas & 1024 & Saturated \\
\hline
\multirow{2}{*}{2} & \multirow{2}{*}{2024 Feb 18} & \multirow{2}{*}{$K_s$} & 5 & 1 & 3 & 600mas & 1024 & Unsaturated \\
{} & {} & {} & 60 & 1 & 5 & 600mas & 1024 & Saturated \\
\hline
\multirow{2}{*}{3} & \multirow{2}{*}{2024 Oct 26} &\multirow{2}{*}{$K_s$}& 0.011 & 100 & 6 & none & 128 & Unsaturated\\
{} & {} & {} & 0.17 & 100 & 5 & none & 1024 & Saturated\\
\hline
\multirow{2}{*}{4} & \multirow{2}{*}{2025 Oct 09} &\multirow{2}{*}{$J$}& 0.011 & 90 & 6 & none & 152 & Unsaturated\\
{} & {} & {} & 60 & 1 & 11 & none & 1024 & Saturated\\
\hline
\hline
\enddata
\end{deluxetable}

\subsection{Data Reduction}
The standard astronomical data reduction process was followed for all observations of HIP 17453 A. First, each raw image was dark subtracted and flat-field corrected, and a bad pixel map was generated from the median darks and flats to remove any bad pixels from the NIRC2 detector. Then, the 2013 first epoch data was distortion corrected using the solution described in \citet{yelda2010}.  A new distortion solution was generated in 2015 due to realignment of the NIRC2 camera; all follow-up data taken in 2024 was distortion corrected using this new solution as described by \citet{service2016}. 

The final step in the data reduction process is point spread function (PSF) subtraction. To remove the stellar PSF we utilize the reference differential imaging (RDI) technique along with Karhunen-Loève Image Processing \citep[KLIP;][]{soummer2012}. With RDI, a reference library of stellar PSFs that are uncontaminated by candidate companions or background stars is constructed from images of single stars across each night of observations. These single stars consisted of other targets in the C-BASS survey. A KLIP algorithm \citep[pyKLIP;][]{pyklip} uses principal component analysis to model a typical PSF for that observation night using the PSFs in the reference library. First, the images are divided into search areas based on user-defined annuli and subsections, and the average value is subtracted from each search area so each one has a mean of zero. Then, the Karhunen-Loève (KL) transform of the reference library PSF model is computed based on the amount of KL modes defined by the user, and an estimate of the PSF of each science image is determined by projecting the image onto the KL transform. Finally, the estimated science image PSF is subtracted from the original science image, producing final images that are largely free of the original stellar PSF \citep{soummer2012, pyklip}.

\begin{figure}
\gridline{\fig{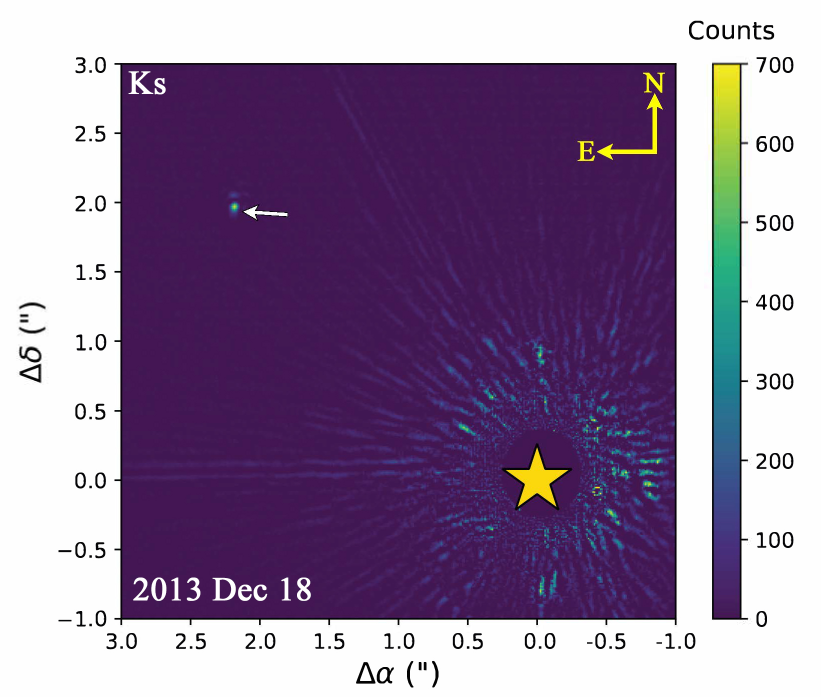}{0.46\textwidth}{Epoch 1}
              \fig{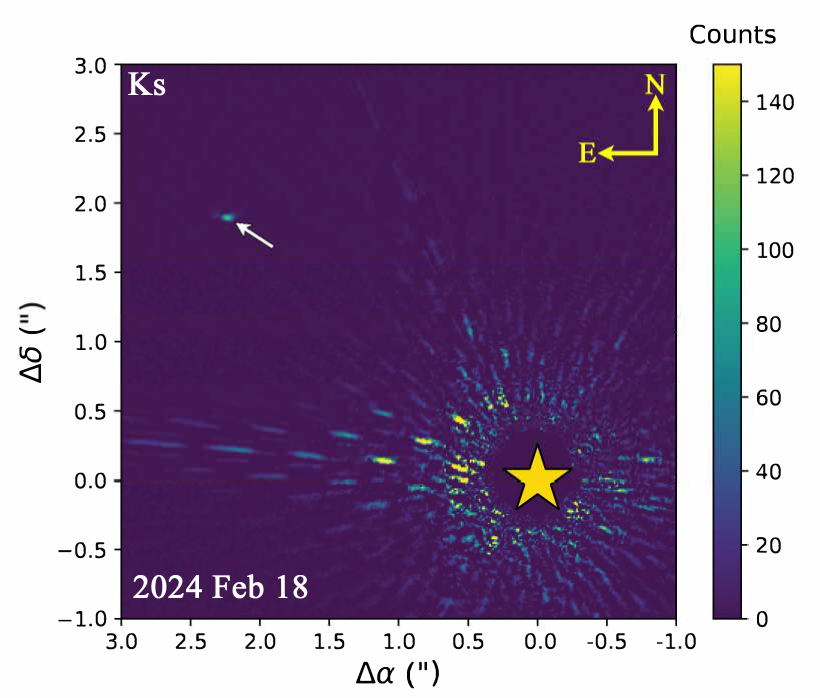}{0.46\textwidth}{Epoch 2}}
\vspace{-2mm}
\gridline{\fig{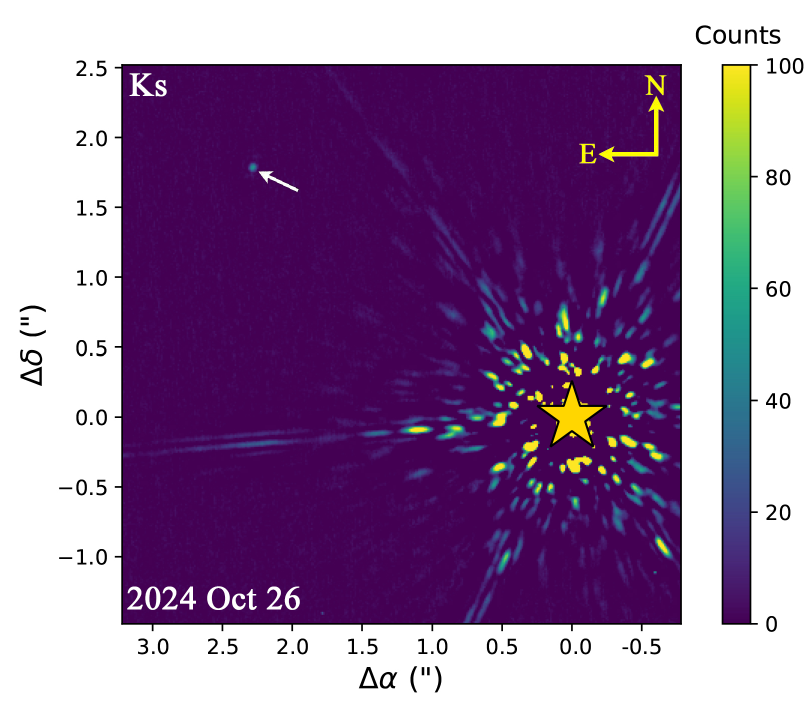}{0.48\textwidth}{Epoch 3}
              \fig{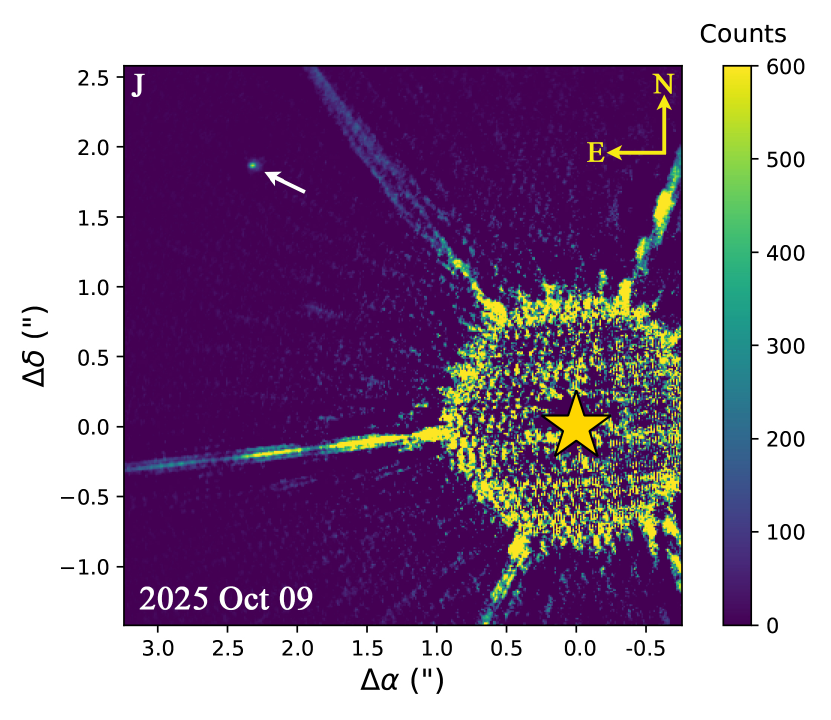}{0.48\textwidth}{Epoch 4}}
    \caption{Keck/NIRC2 $K_S$-band observations of the HIP 17453 system obtained on 2013 Dec 18 (top left), 2024 Feb 18 (top right), and 2024 Oct 26 (bottom left), as well as a $J$-band observation obtained on 2025 Oct 09, showing the location of the primary HIP~17453~A (yellow star) and the substellar companion HIP~17453~B (indicated with white arrow). In the two C-BASS observations (top row), the primary was blocked with a 600mas coronagraphic spot to reduce scattered light and enhance contrast. The top two images have been processed using pyKLIP \citep{pyklip} to subtract the stellar PSF from the image. HIP 17453 B is clearly detected in each observation.} 
    \label{fig:17453obs}
\end{figure}

\begin{figure}
\gridline{\fig{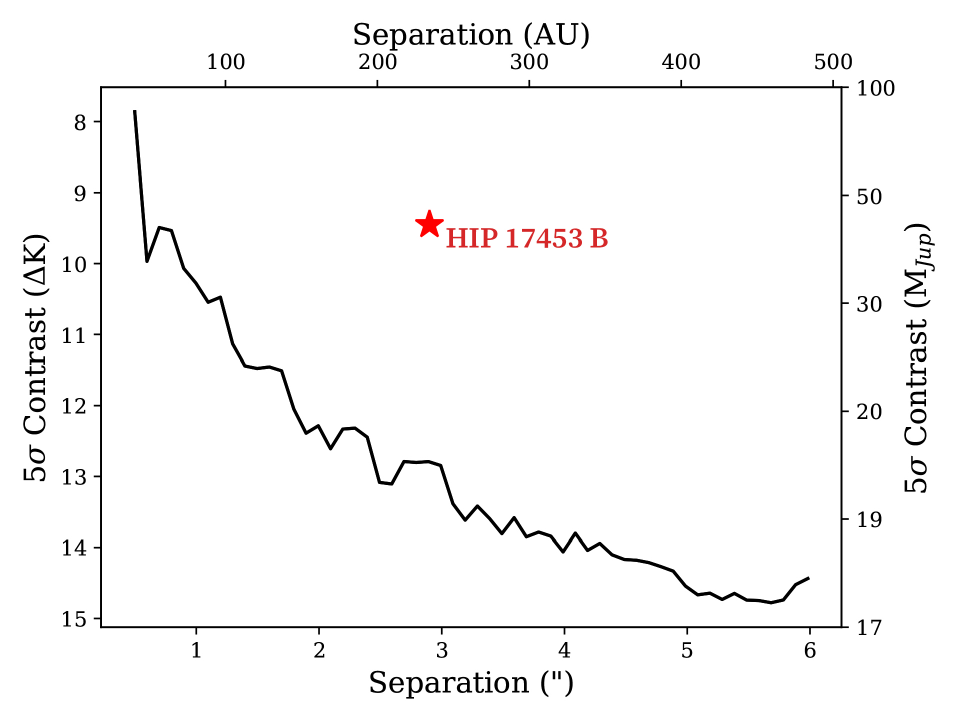}{0.4835\textwidth}{}}
\vspace{-1cm}
\caption{The contrast curve for HIP 17453 A from the initial C-BASS discovery epoch shows even these 5-minute snapshot observations go deep enough to detect brown dwarf companions beyond $\sim$1$^{\prime\prime}$. HIP 17453 B (red star) is detected at high significance in this dataset. The {Sonora Diamondback \citep{sonoradiamondback}} grid of atmosphere and evolutionary models were used to convert $\Delta K_s$ contrast to mass.} 
    \label{fig:contrast_curve}
    \end{figure}

The RDI+KLIP method of PSF subtraction was employed for observations taken during the first two epochs (18 Dec 2013 and 18 Feb 2024), as the observational strategy of the C-BASS survey was structured to create a reference star library for the purposes of RDI. To adequately process the observations and reduce the impact of the stellar PSF on the observations, the final pyKLIP parameters for the reduction of the first two epochs were 32 annuli and 64 subsections. The number of KL modes was defined as 1, 5, 10, 20, and 40, producing a final data cube for each target with all defined KL modes. The 20 KL-mode PSF-subtracted images of HIP 17453 taken over the course of the C-BASS survey are shown in the top panel of Figure 2. For the two observational epochs in Oct 2024, PSF subtraction was instead performed using unsharp masking due to a lack of a reference star library to build the stellar PSF model. These observations are shown in the bottom panel of Figure \ref{fig:17453obs}. 

\subsection{Results: Astrometry }\label{sec:astro}

From a visual inspection of the observations of HIP 17453 A, a candidate companion was identified at an angular separation of $\sim$3$^{\prime\prime}$ in each epoch, as shown in Figure \ref{fig:17453obs}. However, the second-epoch (2024 Feb 17 UT) observations of HIP 17453 A were affected by an instrument rotator issue that impacted the entire night of observations. The NIRC2 rotator was not set back to its original position before the night of observations began, leading to a systematic offset from true North for each image that was taken. However, acquisition images for the AO camera were taken upon slewing to each new target and were unaffected by the NIRC2 rotator issue. Using these acquisition images we were able to determine the systematic offset was equivalent to $\sim$11.2$^{\circ}$ East of North, and applied this correction to the Feb 2024 observations of HIP 17453 A.

The separations and position angles between the primary and companion in each observation were calculated using a Moffat function to determine the relative central pixel positions of each component of the system. These positions were then converted to separation and position angle using the plate scale solution and angle of true north correction as described for NIRC2 by \citet{yelda2010} and \citet{service2016}. These properties are listed in Table \ref{tab:compprop}. 

\begin{figure}
\gridline{\fig{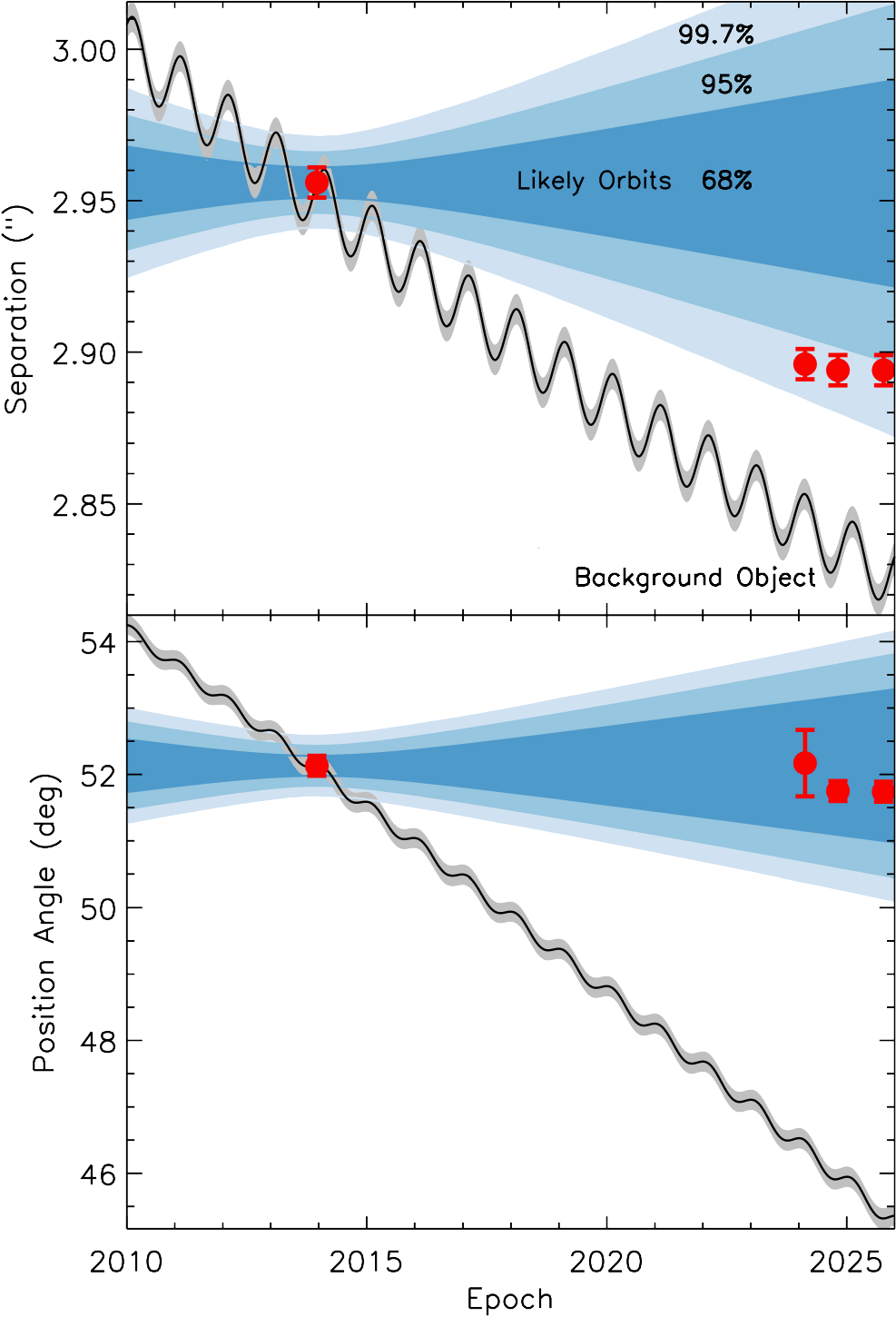}{0.4835\textwidth}{}}
\vspace{-0.5cm}
\caption{{The positions of the candidate companion HIP 17453 B relative to HIP 17453 A in terms of separation and position angle, as measured by Keck/NIRC2 (red points) from 2015–2025. Also shown are the expected range of positions for the likely orbits of a bound companion (blue bands) and a stationary background star (black and gray bands).} HIP 17453 B is clearly bound to HIP 17453 A in terms of separation and position angle.} 
    \label{fig:CPM}
    \end{figure}

In order to determine whether a candidate is a co-moving companion or a background object, the proper motion of the candidate is analyzed with respect to the primary \citep{nici2,drs15,nielsen17}. The relative astrometry over the $\sim$10 year baseline between observations of HIP 17453 B is shown in Fig. \ref{fig:CPM}, with the separations and position angles plotted over time for HIP 17453 B with respect to HIP 17453 A. An object that falls within the gray shaded region is a background star, and objects within the blue shaded region fall within the orbital cone and have bound orbits to the primary. It is highly likely that HIP 17453 B is a bound companion, with astrometry $\sim$15$\sigma$ away from the background track in position angle. No other objects were detected in the observations of HIP 17453. 

{Given that this method assumed that the companion is a stationary background object, we also address the possibility that HIP 17453 B is an unrelated field L dwarf with a non-zero proper motion. We use the empirical $M_J$ relation from \citet{filippazzo2015} and an adopted spectral type (see Section \ref{sec:spty}) for the companion to estimate $M_J$. Combined with an apparent $J$-band magnitude for the companion (see Section \ref{sec:phot}), we calculate a spectrophotometric distance of $75^{+21}_{-17}$ pc for the companion. This is in agreement with the $Gaia$ distance of $80.50^{+0.35}_{-0.29}$ pc for HIP 17453 A and supports HIP 17453 B being a bound companion.}

\subsection{Results: Photometry}\label{sec:phot}

{For the purposes of the C-BASS survey extreme precision photometry was not required; we instead calculate the relative photometry between the primary and companion.} As discussed in Section \ref{subsec:nirc2obs}, both short and long exposures were taken of each C-BASS target including HIP 17453 A. The primary stars are saturated in the long exposures which are required for the sensitivity necessary to image substellar companions, so it is not possible to measure the magnitude difference between HIP 17453 A and B in a single observation. The flux of the primary stars are measured in the unsaturated short exposures, while the long exposures are used to measure the flux of any companions. The flux of the primary is then scaled to the ratio of the saturated and unsaturated exposure times, and the magnitude difference between the primary and companion and the apparent and absolute magnitude of the companion can be calculated. 

In the original strategy of the survey for observations taken during Epochs 1 and 2, the unsaturated exposures of the primary stars were taken with a semi-transparent coronagraphic spot with a measured transmission of $\sim$0.22\% in $K_s$-band \citep{bowler2015}. However, the photometry measured from observations of the star using the coronagraphic spot were relatively variable, with a difference in $\Delta K_s$ of 0.4 mag. {To ensure accurate measurements of the brightness of the central star in order to calculate the relative photometry} of the HIP 17453 system, further follow-up observations were taken on 2024 Oct 26 in $K_s$-band and 2025 Oct 09 in $J$-band. For these observations, the primary was observed unocculted by the coronagraph, while using small subarrays on the detector to reduce the minimum integration time and avoid saturation. Longer exposures were taken using the full detector frame to capture the companion. In all of these cases, the flux from the primary and companion were measured using aperture photometry. The annulus apertures used for both the primary and companion were defined by an inner annulus with a radius extending 10 pixels out from the central pixel of the object, and a 10 pixel-wide aperture encircling the inner annulus to measure the sky background. The measured fluxes were converted to the magnitude difference between primary and secondary, then into apparent and absolute magnitudes for the companion {using the 2MASS-measured magnitudes for the primary}. The flux errors were calculated using apertures to measure the background and standard deviation at equivalent separations to HIP 17453 B at various angles around the central star. These properties are listed in Table \ref{tab:compprop} for HIP 17453 B. 

\begin{deluxetable}{ccccc}
\tablecaption{Properties of HIP 17453 B \label{tab:compprop}}
\tablenum{3}
\tablehead{
\colhead{} &  \colhead{2013 Dec 18} &  \colhead{2024 Feb 18} & \colhead{2024 Oct 26} & \colhead{2025 Oct 09}\\\\[-0.7cm]
\colhead{} &  \colhead{Epoch 1} &  \colhead{Epoch 2} & \colhead{Epoch 3} & \colhead{Epoch 4}
}
\startdata
Separation (arcsec) & 2.956 $\pm$ 0.005 & 2.896 $\pm$ 0.005 & 2.894 $\pm$ 0.005 & 2.882 $\pm$ 0.005\\
Position Angle ($^{\circ}$ E of N) & 52.13 $\pm$ 0.15 & 52.17 $\pm$ 0.5 & 51.75 $\pm$ 0.15 & 51.74 $\pm$ 0.18 \\
$\Delta K_s$ (mag)  & 9.17 $\pm$ 0.50 & 8.77 $\pm$ 0.65 & 9.45 $\pm$ 0.06 & - \\
$m_{Ks}$ (mag)  & 15.20 $\pm$ 0.50 & 14.80 $\pm$ 0.65 & 15.48 $\pm$ 0.06 & - \\
$M_{Ks}$ (mag)  & 10.67 $\pm$ 0.50 & 10.27 $\pm$ 0.65 & 10.95 $\pm$ 0.06 & - \\
$\Delta J$ (mag)  & - & - & - & 10.72 $\pm$ 0.11 \\
$m_J$ (mag) & - & - & - & 16.72 $\pm$ 0.11 \\
$M_J$ (mag)  & - & - & - & 12.19 $\pm$ 0.11
\enddata
\end{deluxetable}

\section{Spectroscopy}\label{sec:spec}
\subsection{Spectroscopic Observations} \label{subsec:gnirsobs}

Following the confirmation that the candidate to HIP 17453 A is a co-moving companion, {and since the photometry is consistent with a brown dwarf given the age of the primary}, we obtained spectra of the system using the Gemini Near Infra-Red Spectrograph (GNIRS) on Gemini-North to confirm the nature of the companion.
We observed the HIP 17453 system with Gemini-North/GNIRS on three dates in 2024: 2024 August 27, 2024 November 28, and 2024 December 21 {(program ID: GN-2024B-DD-102)}. We obtained R$\sim$1800 spectra in cross-dispersed (XD) mode configured with the short blue camera and the 32 lines/mm grating. This setup provided simultaneous wavelength coverage over the 0.85–2.5$\mu$m range with 6 spectral orders, each with 0.15$^{\prime\prime}$ pixel scale. Since HIP 17453 A is an A0 star, it was used as the telluric calibrator and observed immediately after the companion for this purpose. Given the brightness of the primary star, it was observed for 4 seconds per exposure in an ABBA nodding pattern for a total of 16 seconds; each exposure consisted of 2 coadds of 2 seconds each. We obtained the standard GNIRS calibrations, including an argon lamp for wavelength calibration, internal flat fields, and pinhole images to trace and rectify the spectral orders. 

Given the high contrast ratio ($\sim$10 magnitudes) and relatively close separation ($\sim$3$^{\prime\prime}$) of the HIP 17453 system, special observational techniques and data reduction were required to acquire the spectrum of the secondary. The ALTtitude conjugate Adaptive optics for the InfraRed (ALTAIR) adaptive optics system on Gemini-North was used to minimize the amount of light from the primary that reached the position of the companion. The slit (7$^{\prime\prime}$ in length) was oriented orthogonal to the line connecting the primary to the secondary to further reduce light from the primary. {The airmass of the observations remained around 1.00 $\pm$ 0.02 throughout the night, which indicates that wavelength-dependent slit loss should not influence the extracted spectrum despite the slit not being aligned to the parallactic angle.} The companion was nodded along the slit (maintaining the perpendicular orientation of the slit) in an ABBA pattern with an exposure time of 300 seconds per nod, for a total of four exposures.

During the first two GNIRS epochs, a diffraction spike from the star passed over the companion, and so the companion's signal was heavily contaminated by light from the primary star. At the separation of the companion, the flux of the diffraction spike and the flux of the companion were comparable to each other. As the declination of HIP 17453 A (+20:55:43.4) is similar to the latitude of Gemini-North (19.8239$^\circ$), there is little rotation of the diffraction spikes during most of night. On the third epoch, 2024 December 21, the observations were carefully timed to take place close to transit, resulting in observations where the diffraction spike is placed at the maximum possible separation from the companion. Due to the lower stellar contamination of the Dec 21 observation, the analysis only uses this final data set. Figure~\ref{fig:specobspanel} shows the 2D spectrum of HIP 17453 B from a single observation on 2024 Dec 21 with the $J, H$, and $K$ spectral orders visible. The traces of the companion and the diffraction spike are both visible and can be seen to be several pixels separated from each other. At the longest wavelengths in the $K$ order, the companion is significantly brighter than the diffraction spikes, but this reverses when moving to shorter wavelengths.

\subsection{Data Reduction} \label{subsec:gnirsred}

We perform preliminary reduction steps on the data using the Gemini \texttt{IRAF} package (v1.13.1, \citealt{tody1,tody2,giraf}). In the first step, vertical striping, quadrant bias, bad pixels, and cosmic rays are removed. A flat field is generated from observations of the flat lamps, and each frame is divided by that flat field. Observations of the bright primary star are sky subtracted by subtracting the nearest B nod from each A nod (and vice versa). The spectral orders in the 2D images are then rectified so that wavelength increases purely in the vertical direction in the 2D image. The wavelength solution was generated using the spectrum of the argon lamp. We extract the spectrum of the primary star by identifying the peak pixel location as a function of wavelength for each order and computing the median peak location across all wavelengths. The extracted spectrum is the sum of the fluxes from 5 pixels centered on the median peak pixel at each wavelength. This extracted spectrum represents the combination of the primary star and telluric absorption and emission lines. To remove the stellar signal we utilize a PHOENIX synthetic stellar spectrum template \citep{phoenix} with T$_{\rm eff}$ = 9500 K, log(g) = 4.5, and [Fe/H] = 0.0\footnote{Obtained from \url{https://www2.astro.uni-jena.de/Users/theory/for2285-phoenix/grid.php}}, which is then smoothed by a Gaussian to match the resolution of GNIRS. We divide our observed spectrum by the smoothed template, which then gives the atmospheric transmission as a function of wavelength.

\begin{figure}
\gridline{\fig{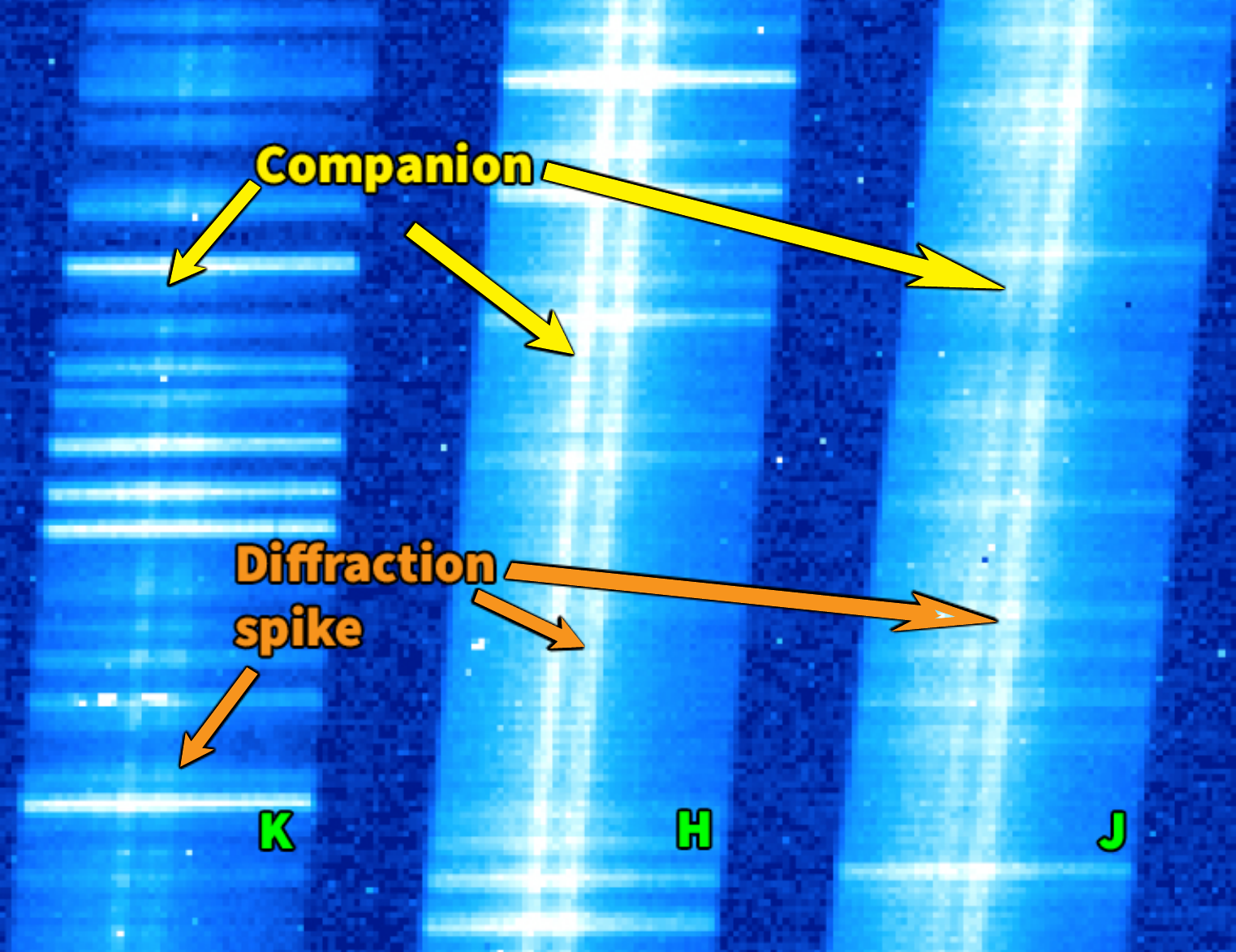}{0.4\textwidth}{}
            \fig{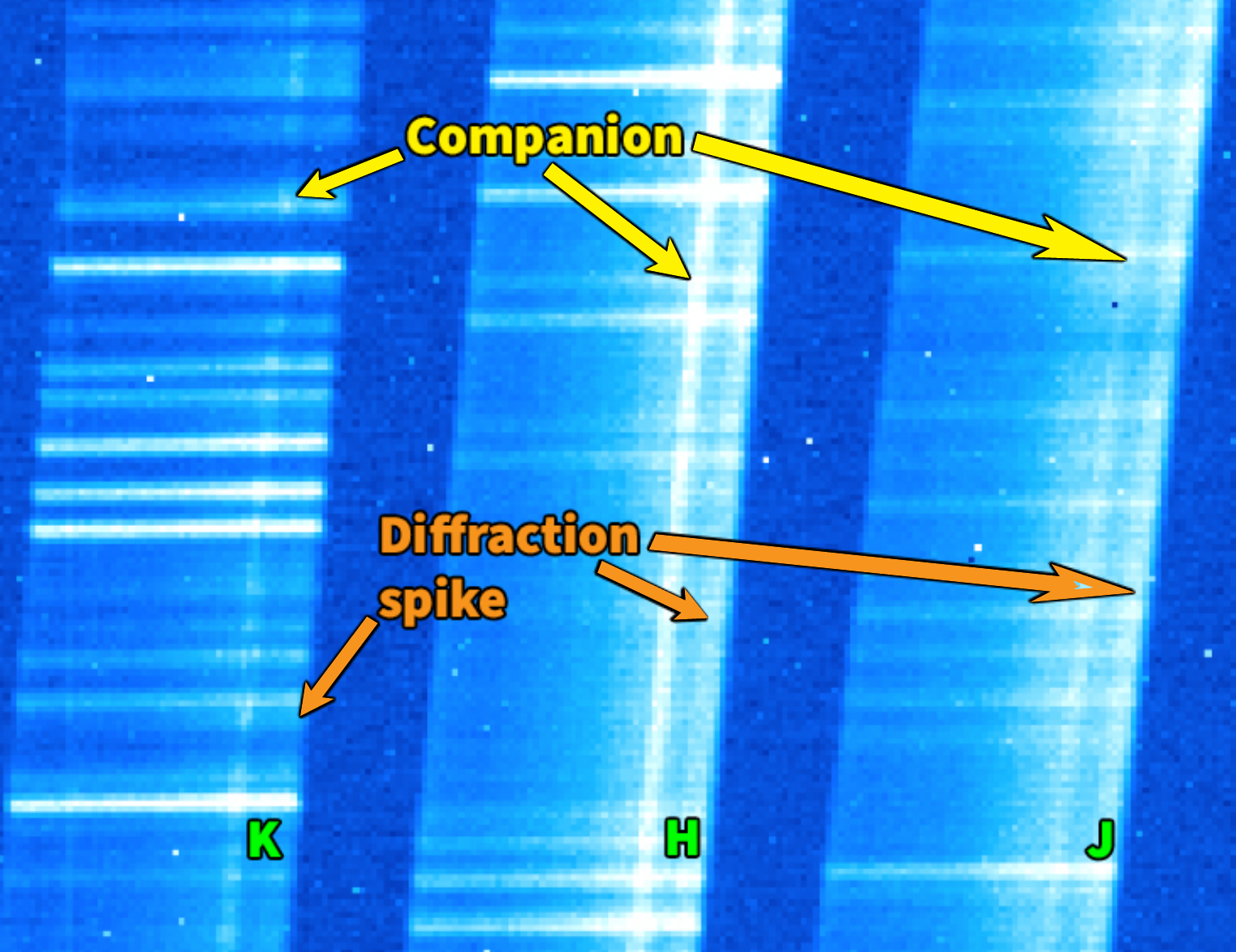}{0.4\textwidth}{}}
\vspace{-1cm}
\gridline{\fig{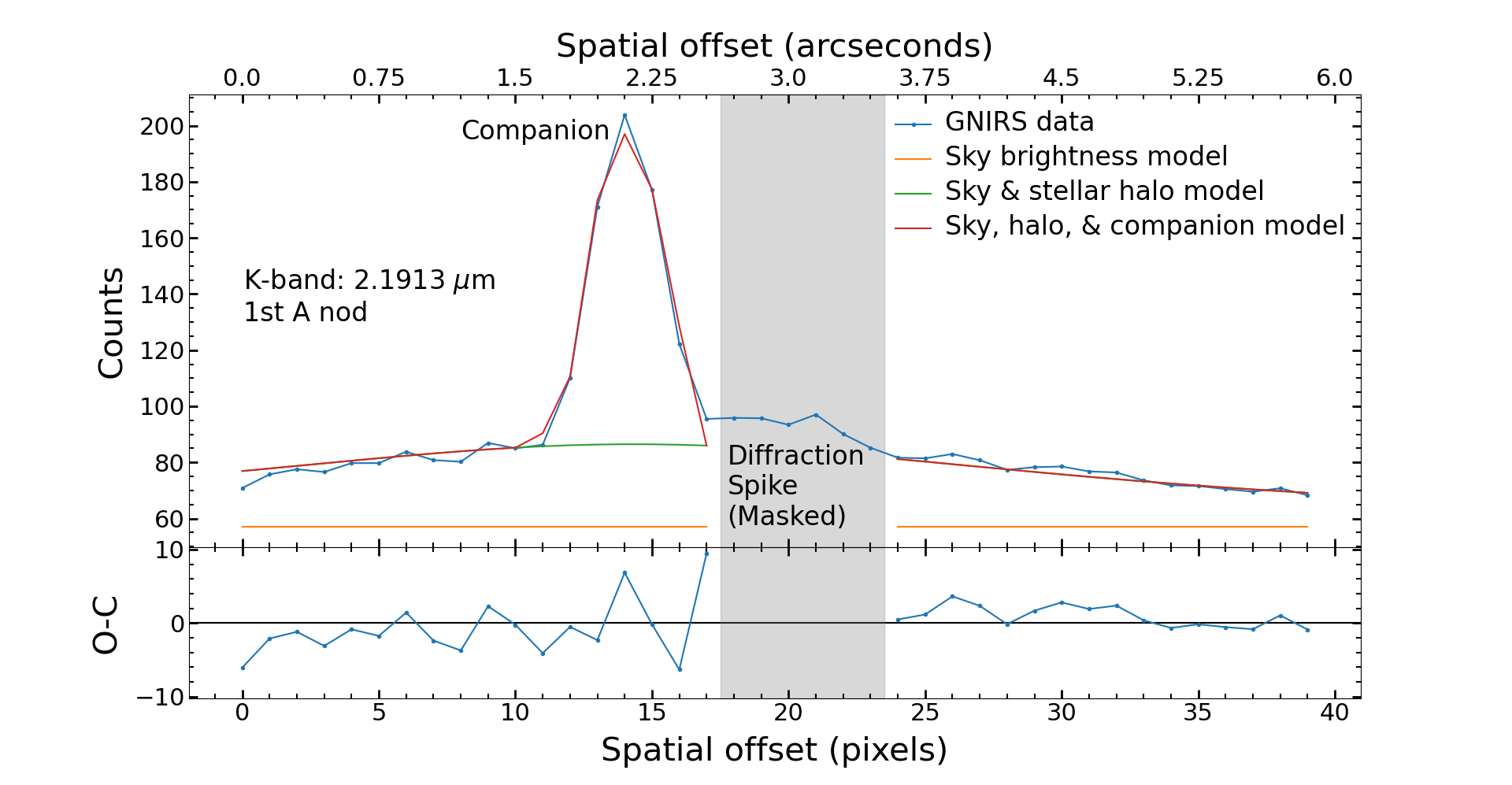}{0.45\textwidth}{}
              \fig{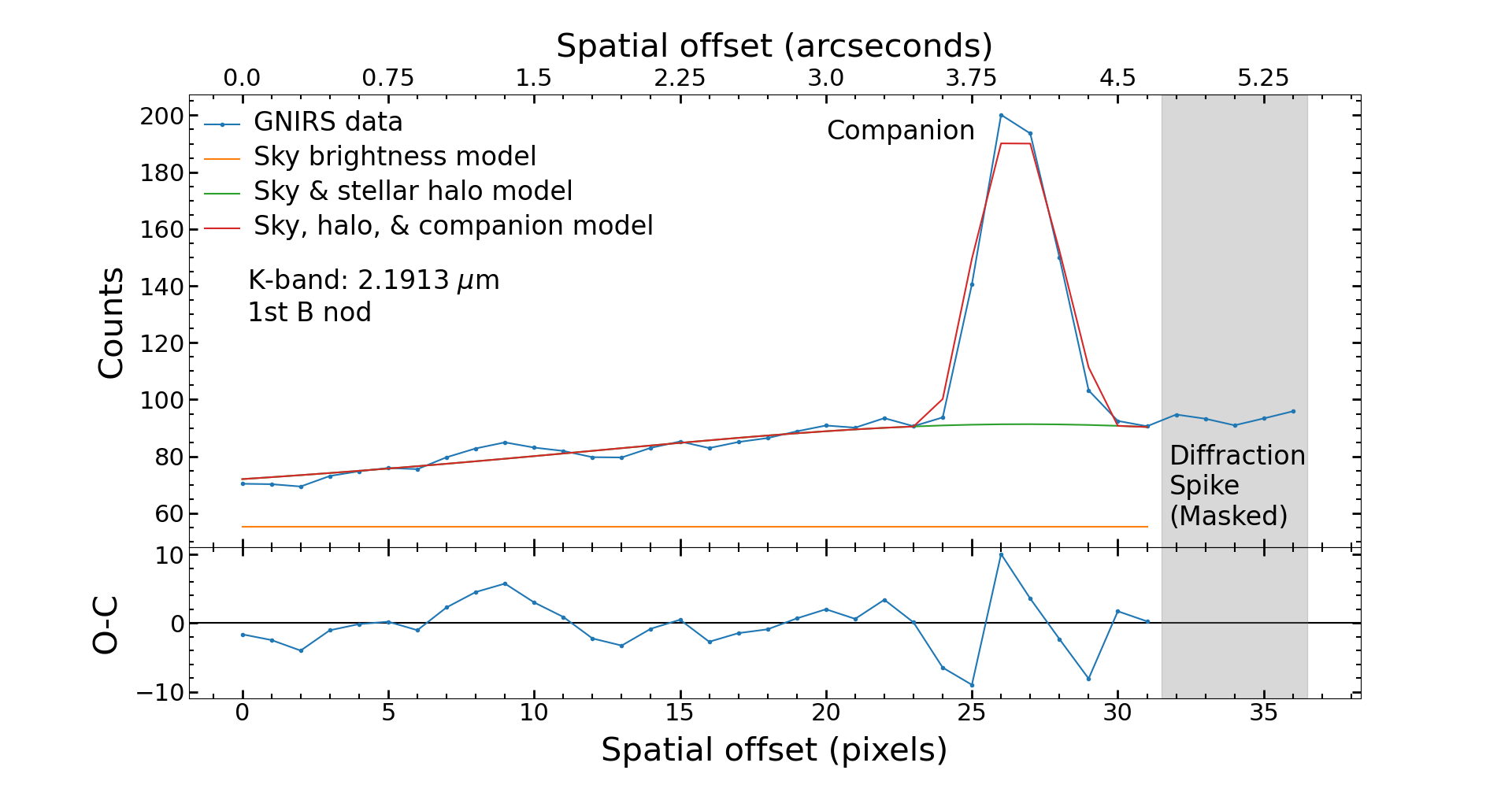}{0.45\textwidth}{}}
\vspace{-1cm}
\gridline{\fig{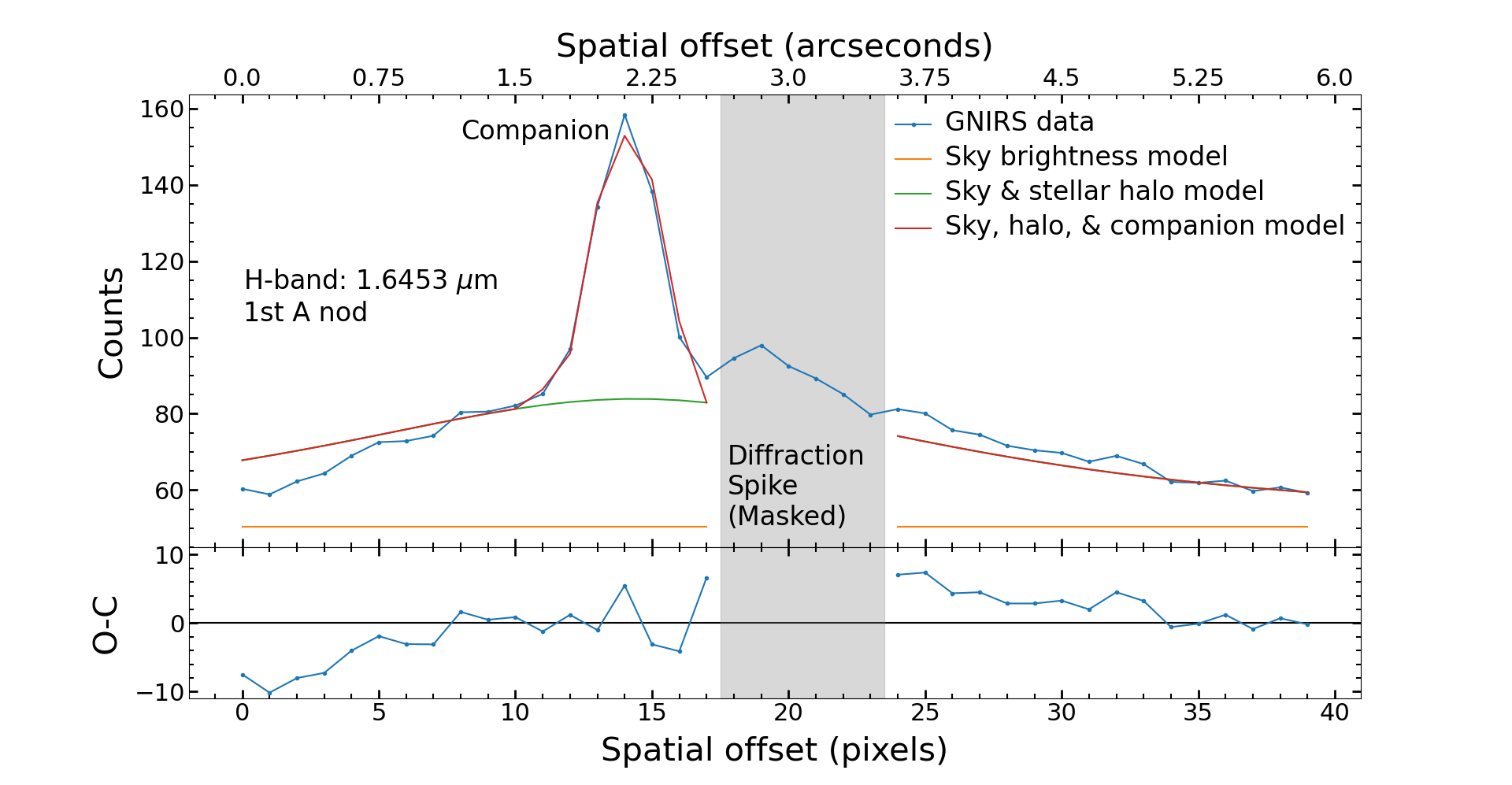}{0.45\textwidth}{}
              \fig{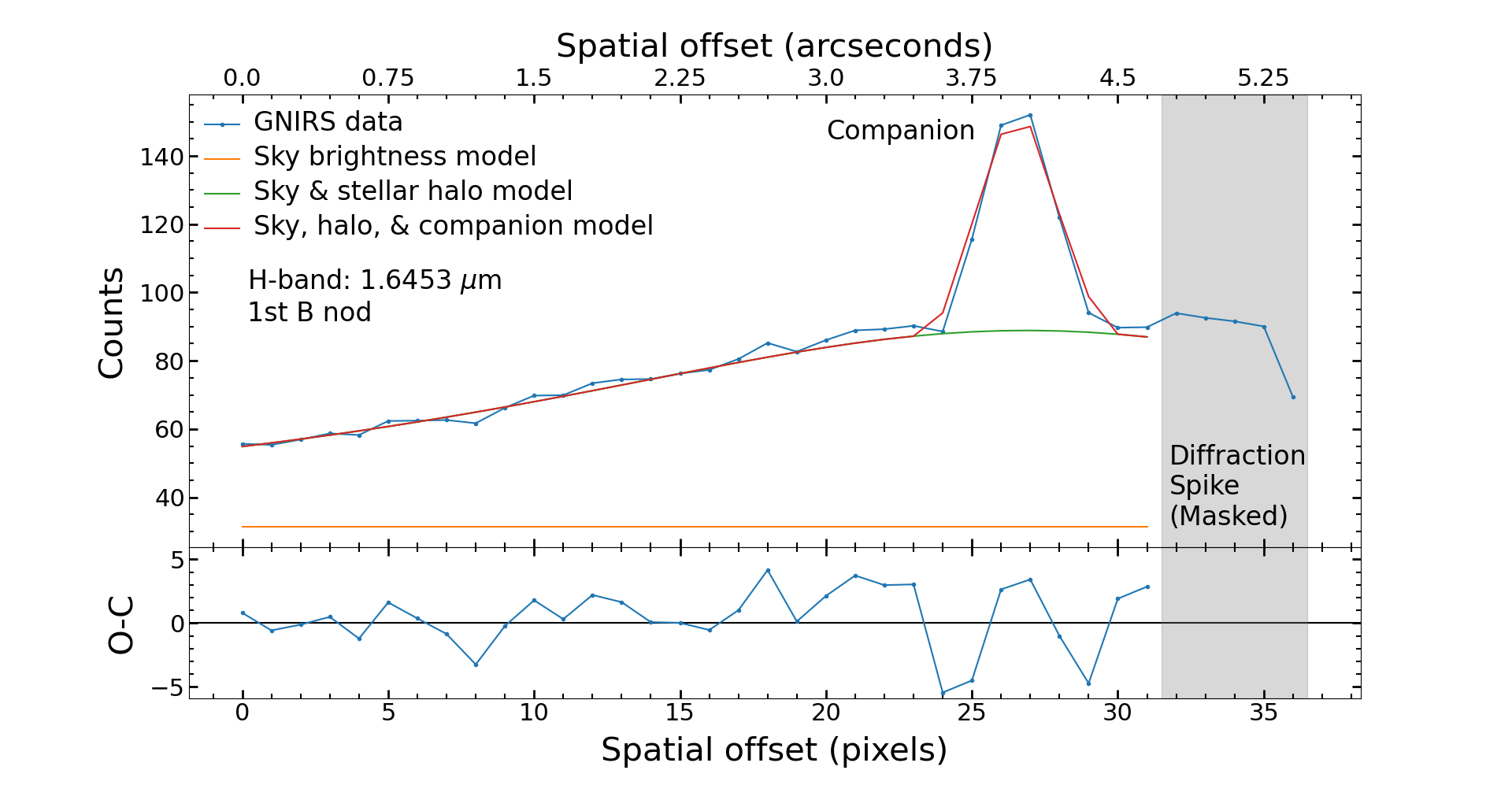}{0.45\textwidth}{}}
\vspace{-1cm}
\gridline{\fig{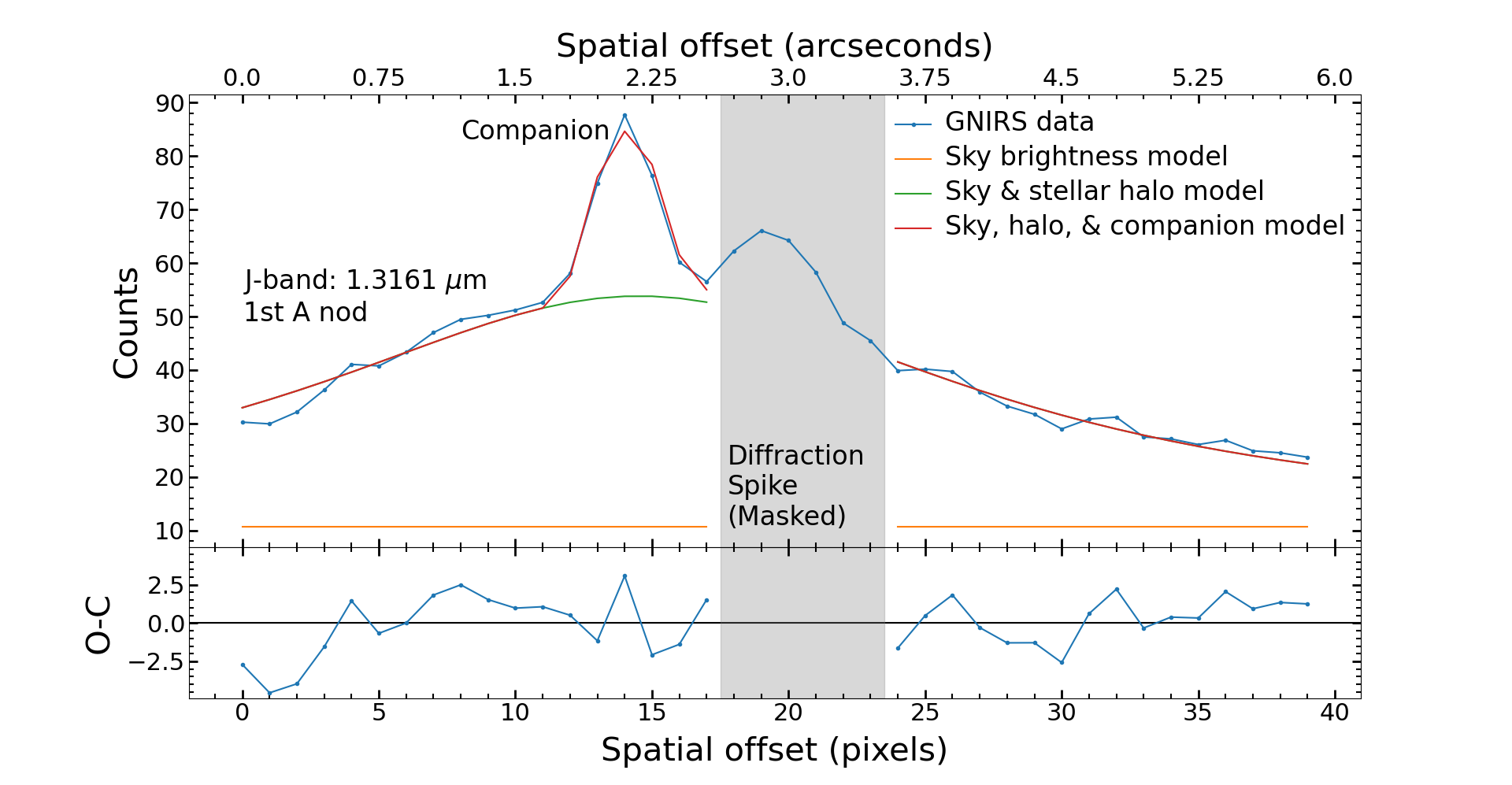}{0.45\textwidth}{}
              \fig{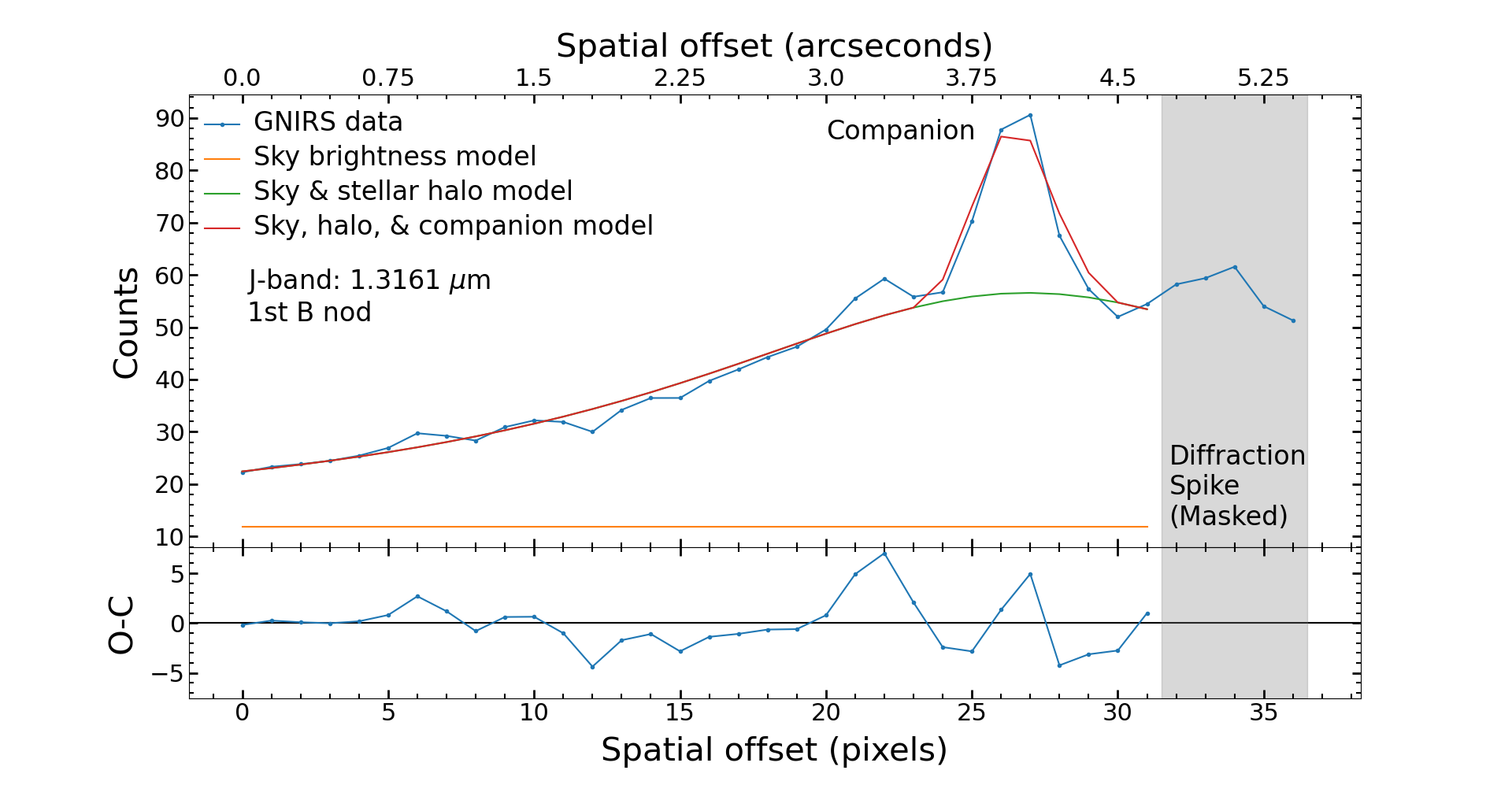}{0.45\textwidth}{}}
    \caption{(\textit{top panel}) Gemini/GNIRS raw spectra of HIP 17453 B from an A-nod (\textit{left}) and B-nod (\textit{right}) exposure on 2024 Dec 21, showing both the companion and a diffraction spike separated by several pixels. Three of the six orders are visible in this view of the GNIRS detector, with $K$ on the far left and $J$ on the right; for each order wavelength is increasing when moving down the chip. (\textit{bottom panels}) Several cuts across single wavelengths for these A-nod and B-nod GNIRS observations of HIP 17453 B, showing flux along the 7$^{\prime\prime}$ slit and the four distinct sources of light present: the cut through the halo of the primary star (the $\sim$40 pixel wide peak across the data), the sky background (a constant offset at all pixels along the slit), the companion flux (the marked $\sim$6 pixel wide peak), and the diffraction spike (marked by grey stripes).}
    \label{fig:specobspanel}
\end{figure}

The extraction of the companion spectrum requires additional steps beyond the standard data reduction steps used for the primary, given the proximity to the primary star and the diffraction spike. Figure \ref{fig:specobspanel} shows flux along the 7$^{\prime\prime}$ slit at a single wavelengths in the $J, H$, and $K$ orders. The data represent four distinct sources of light: the cut through the halo of the primary star (the $\sim$40 pixel wide peak across the slit), the sky background (a constant offset at all pixels along the slit), the companion flux (the $\sim$6 pixel wide peak centered at pixel 14 for the A nod and pixel 27 for the B nod), and the diffraction spike (marked by the grey stripes). We extract the flux from the companion by fitting all of these components except the diffraction spike at each wavelength.

The diffraction spike is masked out with NaNs, with the masked pixels chosen by visual inspection and updated for each of the four exposures since the diffraction spike is quickly rotating near transit. The diffraction spikes in the first A and B nods are symmetrically masked with 5 and 4 pixels respectively, and in the second B nod, the spike is masked with 10 pixels on one side and 5 pixels on the other; finally, the diffraction spike in the last A nod is masked with 5 pixels on one side and 2 on the other. We model the flux $F_\lambda$ across 20 pixels (not including the diffraction spike) centered on the companion location with a six or seven parameter fit. This fit takes the form:

\begin{equation}
    F_\lambda (x) = (S_m x + S_b)  + A L (\mu,\Gamma) + B P(x_c )
\end{equation}

\noindent where x is the pixel number, the first term represents the sky background ($S$) at that wavelength, the second term is a Lorentzian (with median $\mu$, width $\Gamma$, and amplitude $A$) to model the stellar halo, and the third term is a PSF centered on $x_c$ with amplitude $B$ to represent the flux of the companion. The PSF used is the PSF of the primary star extracted previously following AB subtraction at the same wavelength. At most wavelengths the sky background is well represented by a constant value ($S_b$), and we set $S_m = 0$. At some wavelengths near bright telluric emission lines the rectification process (straightening the order on the chip via interpolation) introduces a slope across the slit, and we allow $S_m$ to vary as part of the fitting process for $\sim$10\% of all wavelengths. Bad pixels or cosmic rays were identified by visual inspection and set to NaNs.

A separate fit is carried out at each wavelength with a gradient descent algorithm (Powell's method, \citealt{powell:64}) that minimizes the $\chi^2$ between model and data. Bounds are placed on each parameter to ensure a good fit to the data. 
\begin{itemize}
 \item   Parameters controlling flux levels (sky background $S_m$, Lorentzian amplitude $A$, and the PSF amplitude $B$) are constrained to be between 0 and the maximum measured flux across the pixels of the fit.

    \item Lorentzian width is constrained to be $30 \leq \Gamma \leq 50$
    pixels, which matches the stellar PSF throughout our observations.

    \item When using a non-constant sky background, we restrict $-10 \leq S_b \leq 10$

    \item The center of the Lorentzian ($\mu$) and the center of the PSF ($x_c$) are forced to be the same through the fit. These values, in turn, are constrained to be within 0.5 pixels of a pre-determined center location.
\end{itemize}

The last condition, forcing the center of both the Lorentzian and PSF to be less than a pixel from a given value prevents the fit from exploring unphysical values. The location of this center is chosen by visual inspection of each of the four nods, order by order. This central pixel location is mostly constant across a single order after rectification, but for some orders a correction of 0.5--1.0 pixels is applied partway through the order. We confirm the quality of the fits by visual inspection of every 25th wavelength element. We extract the spectrum by matching the extraction parameters for the star, taking the sum of 5 pixels centered on the peak value of the companion PSF.

\begin{figure}
    \gridline{\fig{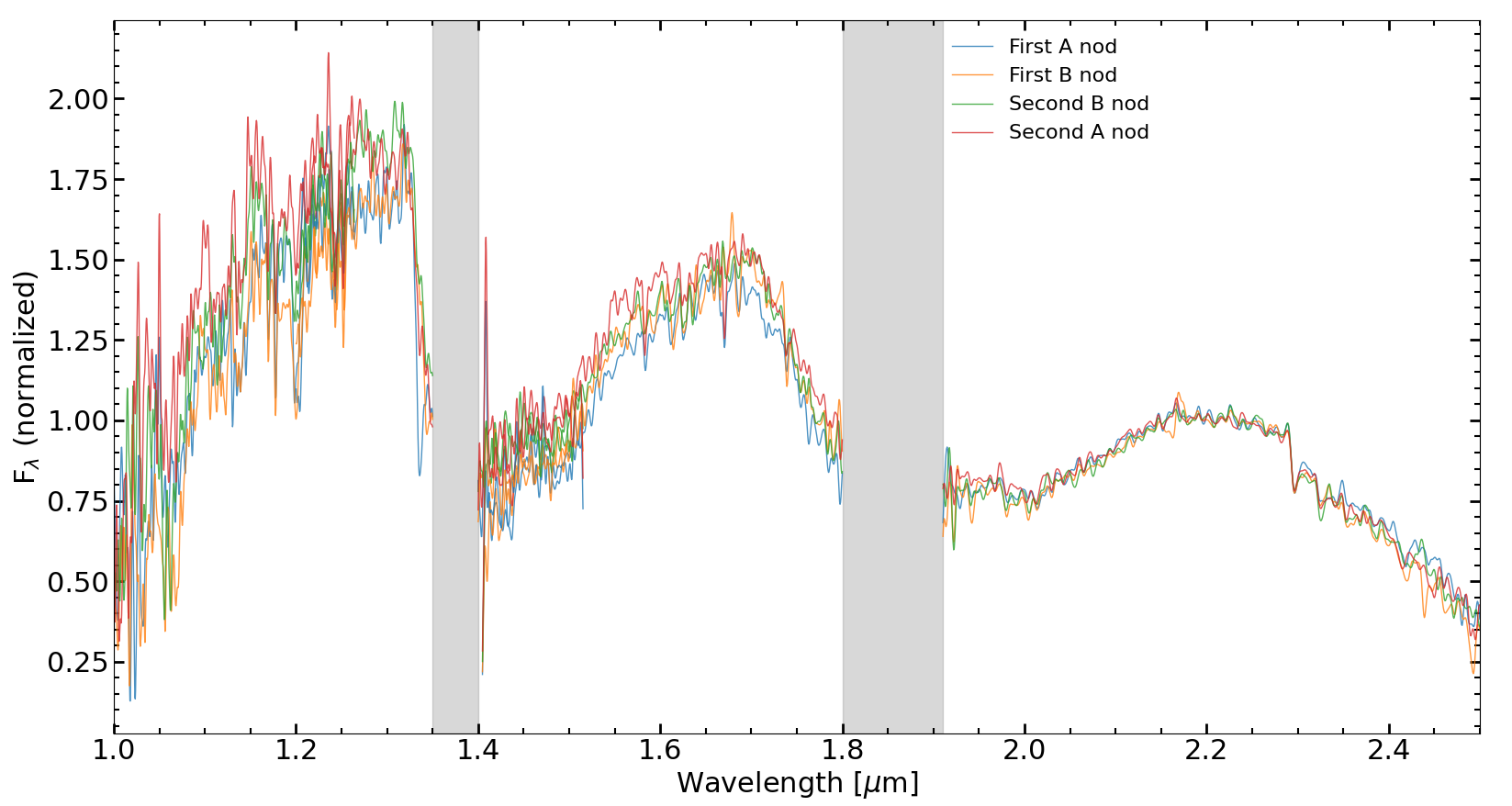}{0.75\textwidth}{}}
    \vspace{-1cm}
    \caption{Comparison of the extracted 2024 Dec 21 GNIRS spectra from each nod position (colored curves) with the average spectrum plotted in black. {The four observations are in good agreement}, with variations on the order of $\leq$10\% excepting the bluest wavelengths in the $J$-band, indicating the success of the reduction method described in $\S$\ref{subsec:gnirsred}.}
    \label{fig:4spec}
    \end{figure}

Figure \ref{fig:4spec} compares the spectra from the 4 nod positions. In addition to the position of the companion changing between the A and B nods, the diffraction spikes move significantly from image to image since these observations were carried out across transit. The four extracted spectra are generally consistent within $\sim$10\%.

\subsection{Results: Spectroscopy}

The shape of the GNIRS spectrum for HIP 17453 B as shown in Figure~\ref{fig:4spec} is consistent with a late-M/early-L spectral type as expected from the age of the primary and the $J$-/$K_s$-band magnitudes of the companion. From an initial visual inspection, the spectrum shows a peak with a modest plateau in the $H$-band (around 1.6 – 1.7 $\mu$m) and prominent CO breaks in the $K_s$-band, which is indicative of a moderately young ($<$500 Myr) substellar object.

{We use the extracted GNIRS spectra for both the primary and companion to confirm our relative $J$- and $K_s$-band photometry, and to calculate relative synthetic $H$-band photometry of the companion. We do not flux calibrate the spectra, as our goal here is to investigate the consistency in the flux ratios between components measured with both NIRC2 and GNIRS. Before the spectra for either object have been corrected for atmospheric transmission, we integrate the spectra over the MKO filter bandpasses to calculate flux ratios for each observation. We then use these flux ratios to calculate $\Delta J$, $\Delta H$, and $\Delta K_s$ magnitudes, calculate the average and standard deviation of the magnitude differences in each bandpass across the four observations, and add a 0.1 mag uncertainty in quadrature to account for possible systematics. {Using this method, we calculate $\Delta J$ = 10.69 $\pm$ 0.11, $\Delta H$ = 9.99 $\pm$ 0.11, $\Delta K_s$ = 9.30 $\pm$ 0.10} and find that these relative synthetic photometry from the GNIRS spectra is generally consistent with the NIRC2 photometry shown in Table \ref{tab:compprop}.}


\section{Discussion}\label{sec:disc}
\subsection{Companion Properties} \label{sec:comp_prop}

\subsubsection{Spectral Type}\label{sec:spty}

\begin{figure}
    \gridline{\fig{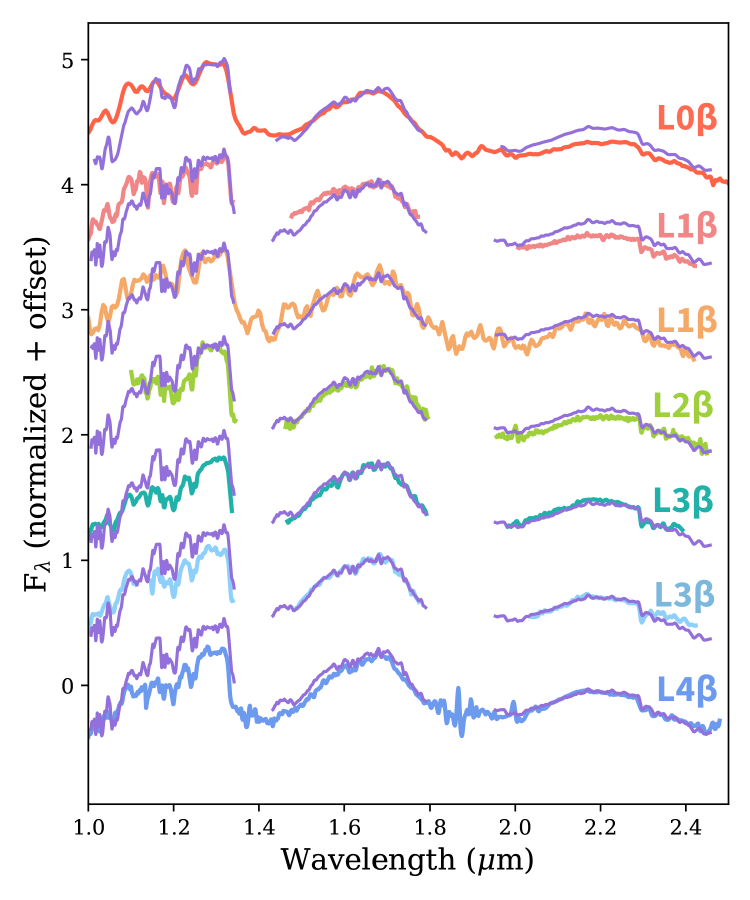}{0.5\textwidth}{}}
    \vspace{-1cm}
\caption{{GNIRS spectrum of HIP 17453 B (purple curves), plotted against L-dwarf near-infrared spectra for the $\beta$ class of objects (ages 100–500 Myr) as defined by \citet{cruz09}  (colored curves) from \citet{bard14} and \citet{gagne15}. All spectra are normalized to the peak H-band flux between 1.67 $\mu m$  $\leq \lambda \leq$ 1.68 $\mu m$. The spectrum of HIP 17453 B is best fit by these early-L spectral standards, with L2 showing the best match across the whole near-IR continuum.}}
    \label{fig:speccomp}
    \end{figure}

{The spectral type of HIP 17543 B was determined through comparison with near-infrared spectra of L-dwarfs. An initial comparison of the GNIRS spectrum shown in Figure \ref{fig:4spec} to the L-dwarf standards from the IRTF Spectral Library \citep{cushing2005,rayner2009} yielded an estimate of $\sim$L2 for the spectral type. However, these are spectral standards for field dwarfs, which are expected to have ages $\gtrsim$1 Gyr (when not members of young moving groups or associations). HIP 17453 A has a markedly younger age ($\sim$280 Myr) than the ages expected for field dwarfs, so we instead determine the spectral type of HIP 17453 B using a sample of L-dwarfs with ages closer to the calculated age of HIP 17453 system. \citet{cruz09} defined a new spectral typing scheme based on the optical spectra of L-dwarfs, where a $\gamma$ classification indicates an age of $\sim$10 Myr and a $\beta$ classification indicates an age of $\sim100$ Myr. Objects in the $\gamma$ class are typically considered ``very low gravity", whereas objects in the $
\beta$ class are considered to have ``intermediate gravity" \citep{kirk10,allers13}. We therefore utilize near-infrared spectra for $\beta$ objects from the Montreal Spectral Library \citep{gagne15,rob16} and \citet{bard14} to compare to the spectrum of HIP 17453 B and determine the spectral type, as shown in Figure \ref{fig:speccomp}. The GNIRS spectrum of HIP 17453 B and the spectra for the $\beta$ objects were binned to the same resolution and normalized to the peak H-band flux (1.67 $\mu m$  $\leq \lambda \leq$ 1.68 $\mu m$). An inspection of the HIP 17453 B spectrum compared to the $\beta$ spectra indicates of a spectral type of {L2 $\pm$ 1}, in particular when comparing the continuum slope in $H$-band and the CO bandheads in $K$-band.} 

\begin{figure}
    \gridline{\fig{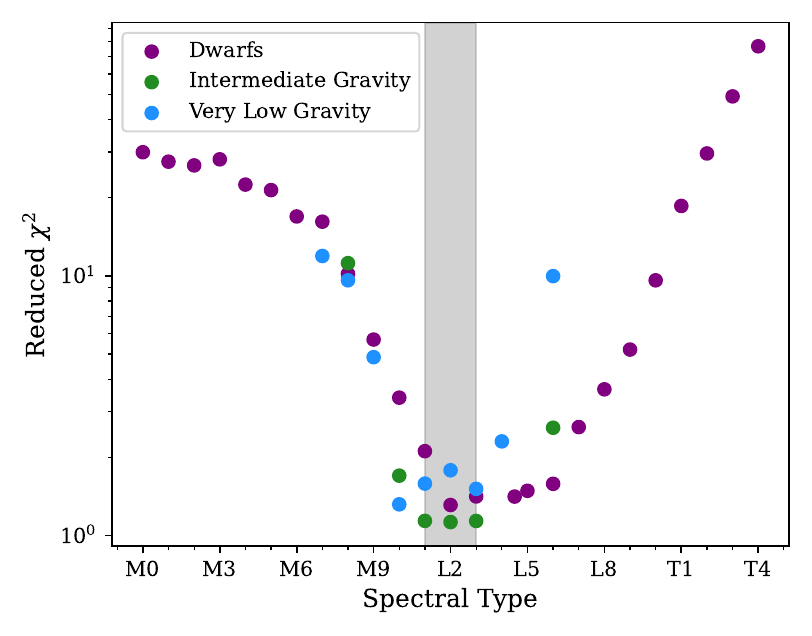}{0.5\textwidth}{}}
    \vspace{-1cm}
\caption{{The reduced $\chi^2$ of the comparison between the reduced GNIRS spectrum of HIP 17453 B and field dwarfs from the IRTF spectral library  \citep{cushing2005,rayner2009} and very low/intermediate gravity objects from the Montreal Spectral Library \citep{gagne15}, with different classes of objects denoted by the various colors. The spectrum of HIP 17453 B is best fit by the intermediate gravity early-L dwarf objects. A spectral type of {L2 $\pm$ 1} is adopted as denoted by the gray shaded region.}}
    \label{fig:chi2}
    \end{figure}

{We verify this designation using \texttt{splat}, the SpeX Prism Library Analysis Toolkit \citep{splat}, to compare the spectrum of HIP 17453 B to the dwarf, intermediate gravity, and very low gravity objects. The GNIRS spectrum of HIP 17453 B showed the best match to the intermediate gravity objects around the L2 range, with the reduced $\chi^2$ metric for this comparison as a function of spectral type shown in Figure \ref{fig:chi2}. The different types of objects are indicated by the various colors, and the {L2 $\pm$ 1} region is shaded in gray. We also apply the spectral and surface gravity index-based classification outlined in \citet{allers13} to determine the spectral type and gravity classification of HIP 17453 B, with the results shown in Table \ref{tab:indices}. The spectral indices used in this analysis are sensitive to spectral type but insensitive to surface gravity, and return a mean spectral type of {L2 $\pm$ 1} for the spectrum of HIP 17453 B. The surface gravity indices can serve as an indicator of youth. The FeH indices are weaker in younger, low-gravity objects than field-gravity objects, while the opposite is true for the VO index; the H-band continuum is more triangular in objects with lower gravity, and the $K$-band continuum of young/low-gravity objects is more positive than in normal field dwarfs \citep{allers13}. The classification of the surface gravity indices depends on the spectral type of the object being analyzed, and for HIP 17453 B and its spectral type of L2, the VO$_z$ index is more consistent with a very low gravity (VLG) object (assigned a gravity score of 2), while the FeH$_J$, K I$_J$ and $H$-cont indices are consistent with intermediate gravity (Int-G) objects (assigned a gravity score of 1). The mean surface gravity score is 1.25, which is consistent with an intermediate-gravity object. Based on these analyses, we adopt a spectral type of {L2 $\pm$ 1} for HIP 17453 B.}

\begin{deluxetable}{lccc}
\tablecaption{Spectral and Surface Gravity Indices for HIP 17453 B \label{tab:indices}}
\tablenum{4}
\tablehead{
\colhead{ } &  \colhead{Reference} &  \colhead{Result} & \colhead{Classification}}
\startdata
\textit{Spectral Indices} & & & \\
\hline
H$_2$O & \citet{allers07} & 1.340 & L4 $\pm$ 0.5 \\
H$_2$O-D & \citet{mclean03} & 0.930 & L2 $\pm$ 1 \\
H$_2$O-1 & \citet{slesnick04} & 0.675 & L1 $\pm$ 1\\
H$_2$O-2 & \citet{slesnick04} & 0.829 & L2 $\pm$ 0.5 \\
FeH & \citet{slesnick04} & 0.762 & L2 $\pm$ 1 \\
 & &  & \textit{L2 $\pm$ 1} \\
 \hline
\textit{Surface Gravity} & & & \\
\textit{Indices} & & & \\
\hline
VO$_z$ & \citet{allers13} & 1.272 & 2 (VLG) \\
FeH$_J$ & \citet{allers13} & 1.140 & 1 (Int-G) \\
K I$_J$ & \citet{allers13} & 1.037 & 1 (Int-G) \\
$H$-cont & \citet{allers13} & 0.906 & 1 (Int-G) \\
 & &  & \textit{1.25 (Int-G)}
\enddata
\end{deluxetable}

\subsubsection{Mass and Effective Temperature}\label{sec:mass+teff}

To estimate the mass of HIP 17453 B, we first use the spectral type, L2 $\pm$ 1, adopted in Section \ref{sec:spty} and the empirical relation for young ($< 500$ Myr) L dwarfs from \citet{filippazzo2015} to calculate a bolometric correction (BC) for $M_{K_s}$. Our $M_{K_s}$ derived from the Epoch 3 NIRC2 photometry has the highest precision and the $M_{K_s}$ BC relation has a lower root mean square (rms) deviation than the $M_{J}$ relation. We propagate the uncertainty from the spectral type through the BC relation. Assuming $M_{bol, \odot} = 4.74$, we then calculate a bolometric luminosity of log $L_{bol} = -3.79 \pm 0.06$ $L_\odot$ for HIP 17453 B. We use $L_{bol}$ and the age of the primary to interpolate the Sonora Diamondback \citep{sonoradiamondback} substellar evolution models in mass and age, assuming solar metallicity. We utilize the hybrid evolution grid, where the models transition from cloudy to cloud-free below the L-T transition with the sedimentation efficiency set to $f_{\rm sed} = 2$. We implemented the MCMC package \textit{zeus} \citep{zeus}, which fits a Bayesian Gaussian Mixture to the walkers to facilitate mode jumping. We use 50 walkers, $1 \times 10^4$ burn-in iterations, and $1 \times 10^5$ final iterations to generate a posterior sample size of $5 \times 10^6$.

\begin{figure}
   \gridline{\fig{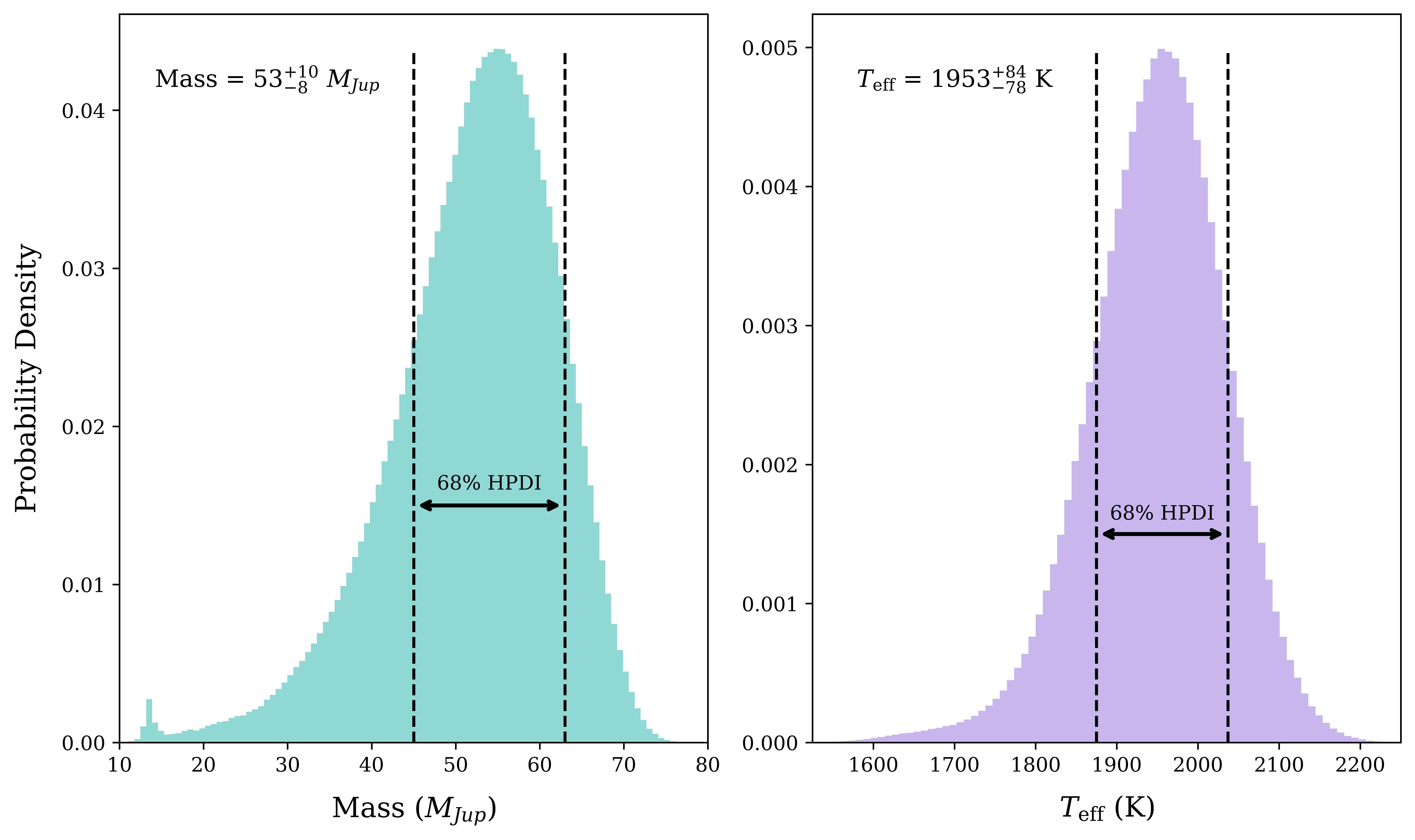}{0.95\textwidth}{}
              }
    \vspace{-1cm}
    \caption{{MCMC mass (\textit{left}) and effective temperature (\textit{right}) posterior samples for HIP 17453 B from interpolation of the Sonora Diamondback \citep[][]{sonoradiamondback} evolution model grid. We calculate the 68\% HPDI of the posteriors to remove the influence of the low-probability peak at young ages and in the low-mass tail of the mass posterior.}}
    \label{fig:masses}
    \end{figure}

Our resulting posterior for the mass of HIP 17453 B is shown in Figure \ref{fig:masses}. The majority of the probability density lies between $\sim 45-65$ $M_{\rm Jup}$, but there is a peak in the low-mass tail near the deuterium burning limit ($\sim 13$ $M_{\rm Jup}$). This peak corresponds to an age of $\sim 10$ Myr, making it unlikely given the age of the primary ($280 \pm 125$ Myr). In order to remove the influence of the low-probability tail from our adopted mass, we calculate the 68\% highest probability density interval (HPDI) of the posterior. The HPDI is the shortest interval that contains a given percentage of the probability mass. We report a final mass of $53^{+10}_{-8}$ $M_{\rm Jup}$ for HIP 17453 B. 

\begin{figure}
    \gridline{\fig{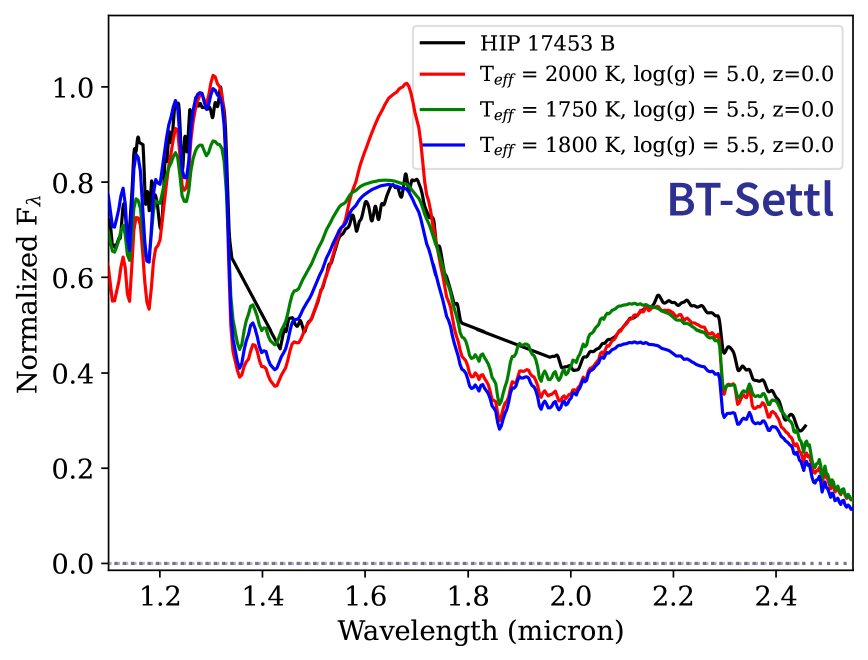}{0.45\textwidth}{}
               \fig{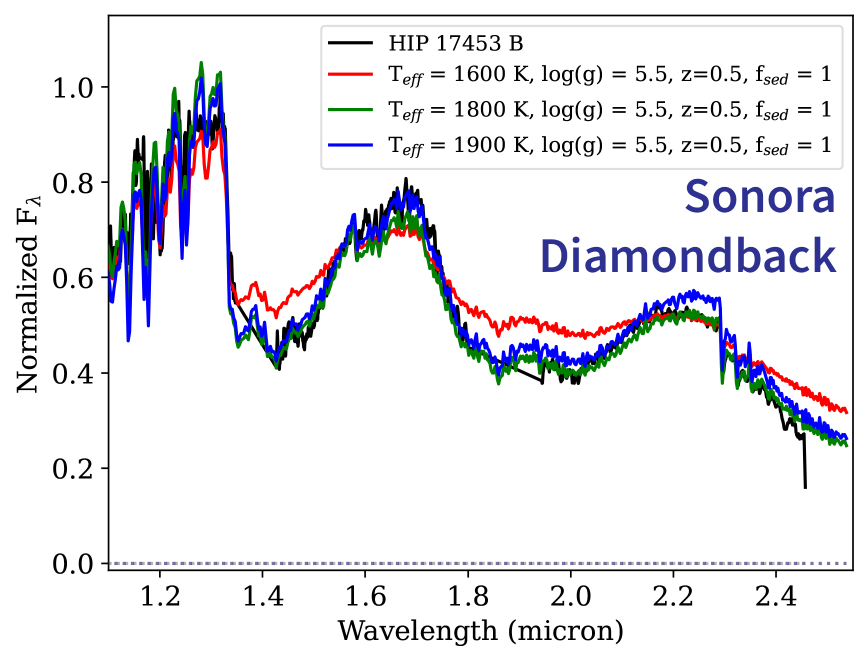}{0.45\textwidth}{}
              }
    \vspace{-0.75cm}
    \caption{Comparison between the HIP 17453 B GNIRS spectrum and the BT-Settl (left) and Sonora Diamondback (right) model grids using \texttt{splat} \citep{splat}. The best-fitting temperatures and surface gravities range between $T_{eff}$ = 1600–2000 K and log $(g)$ = 5.0 – 5.5. The Sonora Diamondback fits prefer cloudy models.} 
    \label{fig:modelcomp}
    \end{figure}

{Using the same methods as above, we calculate the effective temperature of HIP 17453 B to be $1953^{+84}_{-78}$ K.} We also utilize the model-grid fitting feature of \texttt{splat} to compare between the GNIRS spectrum and model spectra. The three best-fit models from the BT-Settl and Sonora Diamondback model grids are shown in Figure \ref{fig:modelcomp}. The best-fitting temperatures and surface gravities range between $T_{\rm eff}$ = 1600–2000 K and log $(g)$ = 5.0 – 5.5. The Sonora Diamondback fits prefer cloudy models over cloud-free models, as expected for the temperature of the companion.

\subsection{Orbit Analysis} \label{sec:orbits}

From the relative astrometry of HIP 17453 B compared to HIP 17453 A, especially apparent in Figure~\ref{fig:CPM}, it is clear that there is orbital motion of HIP 17453 B over the $\sim$10 year observational baseline of the C-BASS observations. While there is little change in position angle over this timespan, the $\sim$3$^{\prime\prime}$ separation decreases by $\sim$50 mas (about 5 NIRC2 pixels). This small amount of motion toward the star is generally consistent with an edge-on orbit ($i \approx 90^\circ$) and a relatively long orbital period ($P \gtrsim 1000$ yr).

\begin{figure}
    \gridline{\fig{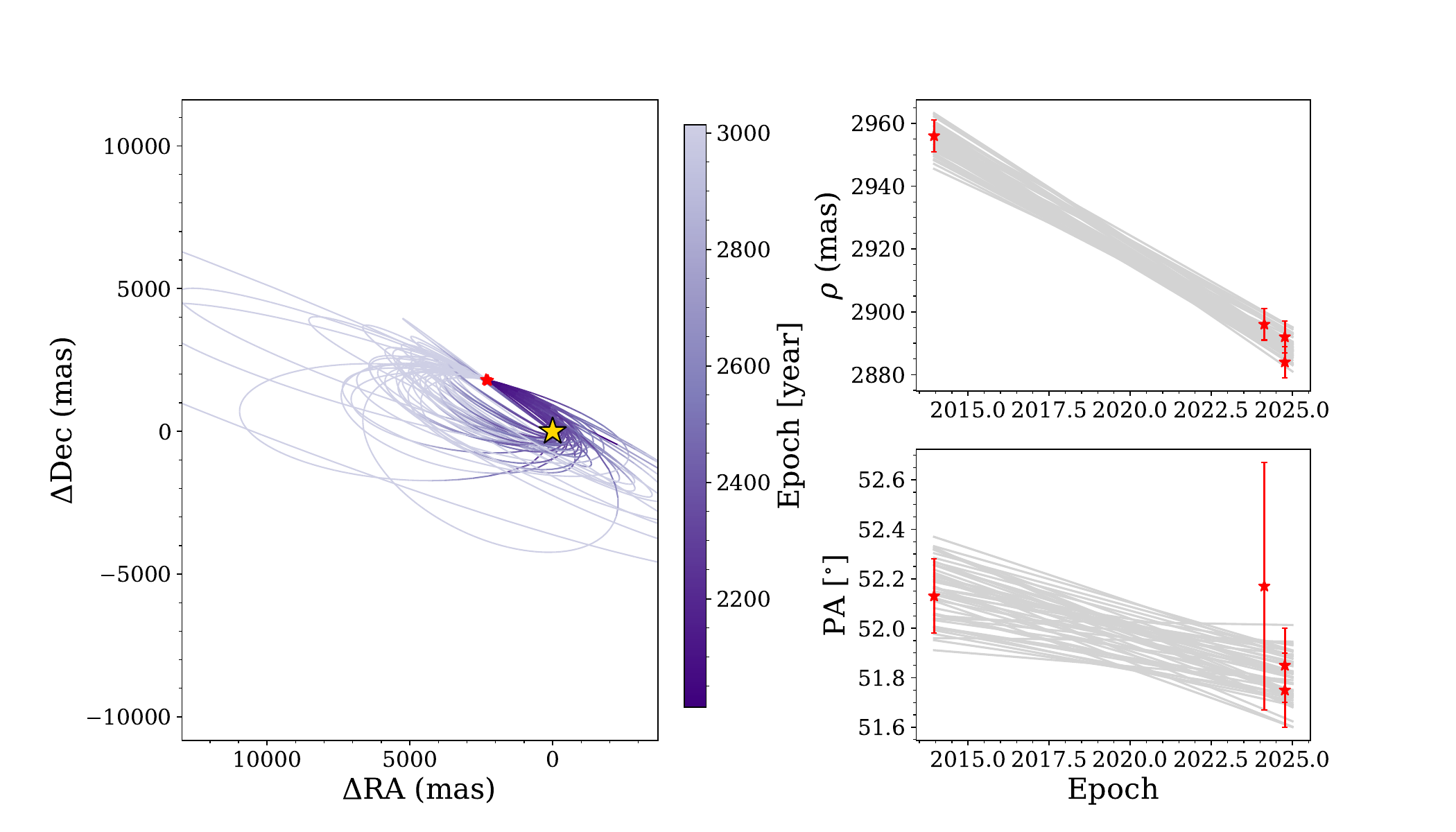}{0.95\textwidth}{}}
        \vspace{-1cm}
    \caption{(\textit{left}) Relative astrometry of HIP 17453 B (red stars) with respect to HIP 17453 A (yellow star), showing 50 randomly selected draws from the orbital posterior (purple ellipses) generated using \textit{orbitize!} \citep{blunt2017,blunt2019}. (\textit{right}) The relative separations (top) and position angles (bottom), with the observed NIRC2 measurements and errors (red points). The orbit is not well constrained from the small amount of orbital motion over the past decade, as expected.}
    \label{fig:H17orb}
    \end{figure}

\begin{figure}
    \gridline{\fig{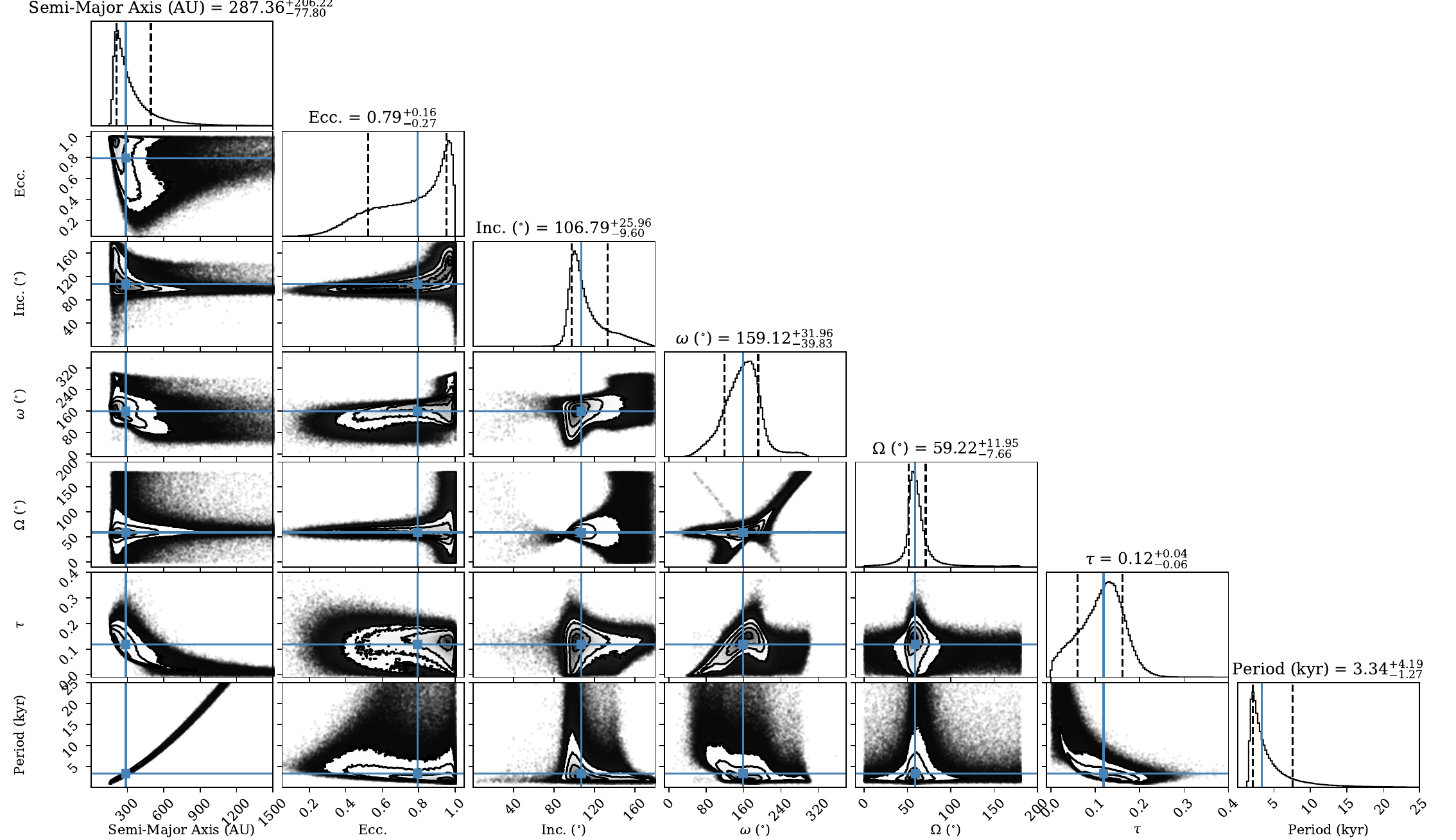}{1.0\textwidth}{}}
        \vspace{-0.75cm}
\caption{Corner plot showing the distributions for the orbital parameters generated by the OFTI orbitize! orbit fitting rejection sampling algorithm. An orbit fit to the motion of HIP 17453 B is generally consistent with a long-period ($P = 3340^{+4190}_{-1270} yr$), edge-on orbit. Since the motion is largely in the separation direction (i.e. toward the star), edge-on orbits are preferred with an inclination angle of $i = 106.8^{+26.0}_{-9.6}{} ^\circ$.}
    \label{fig:H17orbitcorner}
    \end{figure}

To explore the orbital properties of the companion we utilize the OFTI (Orbits For The Impatient, \citealt{blunt2017}) rejection sampling algorithm in \texttt{orbitize!} \citep{blunt2019}. We set priors on the parallax of $\pi$ = 12.3772 $\pm$ 0.049 mas from $Gaia$ DR3 \citep{gaiadr3} and total mass of M$_{tot} = 2.12 \pm 0.1$ M$_\odot$. This total mass comes from propagating errors on the mass of the star ($2.08 \pm 0.09$ M$_\odot$) and the mass of the brown dwarf ($0.043 \pm 0.01$ M$_\odot$). We fit all four relative astrometry measurements with error bars as given in Table~\ref{tab:compprop}. Several random draws from the posterior are shown in Figure \ref{fig:H17orb}, and the corner plot showing the constraints on the orbital parameters is shown in Figure \ref{fig:H17orbitcorner}.

As expected, several orbital parameters are only marginally constrained with the small amount of orbital motion. We find a semi-major axis of $287^{+206}_{-78} AU$, corresponding to an orbital period of $3340^{+4190}_{-1270}$ years. The fit also prefers larger eccentricities of $0.79^{+0.16}_{-0.27}$ and edge-on orbits with an inclination of $106.8^{+26.0}_{-9.6}{} ^\circ$. There is a covariance between inclination angle and eccentricity, with more face-on orbits generally having higher eccentricity, as expected for motion mainly directed toward the star.

The $Hipparcos$-$Gaia$ Catalogue of Accelerations (HGCA, \citealt{brandt2021}) reports a $\chi^2$ value of 3.893 for HIP 17453 A. This statistic represents how well a constant proper motion model fits the $Hipparcos$ and $Gaia$ data for the star, with larger values more consistent with acceleration due to orbital motion. For a $\chi^2$ distribution with two degrees of freedom we would expect a value of 3.893 or larger 14\% of the time. Thus, there is no significant evidence of acceleration from the HGCA, which is not surprising given the smaller mass of the companion and the long orbital period.



\subsection{Comparison with Known Companions}

\begin{figure}
    \gridline{\fig{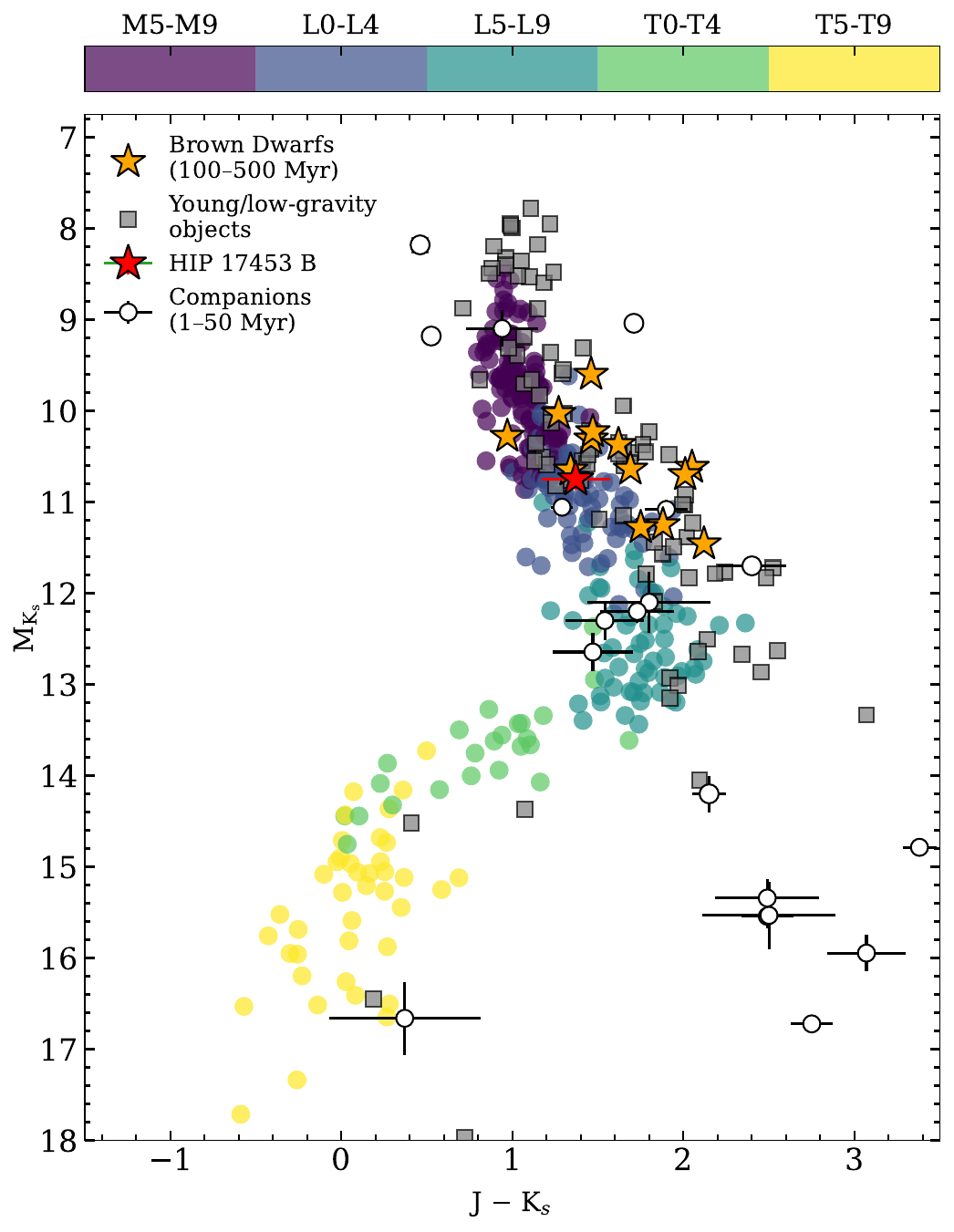}{0.5\textwidth}{}}
        \vspace{-1cm}
    \caption{Color–magnitude diagram showing M$_{Ks}$ vs. J–K$_s$ for various young {and/or} low-mass objects, generated using \textit{species} \citep{stolker2020,species}. On this plot are field objects (color coded by M, L, and T spectral types), young {and/or} low-gravity dwarf objects (gray squares), known directly imaged planetary {or brown dwarf companions (white circles), as well as known directly imaged free-floating brown dwarfs or brown dwarf companions with ages between 100–500 Myr (orange stars). HIP 17453 B highlighted with a red star and is consistent with the 100–500 Myr brown dwarf population}. 
    }
    \label{fig:CMD}
    \end{figure}

The position of an object on a near-IR CMD can be used to provide context for its observed properties and assess its consistency with empirical sequences and theoretical predictions for substellar objects. Figure \ref{fig:CMD} shows HIP 17453 B on an $M_K$$_s$ versus $J - K_s$ CMD compared with known field dwarfs and young or low-gravity objects \citep{best2024}, and various directly-imaged planets and substellar companions, {as well as other L–dwarfs around the same age range (100–500 Myr)}, generated using \textit{species} \citep{stolker2020,species}. This comparison allows us to evaluate the consistency of HIP 17453 B's properties with known substellar populations and evolutionary models. The 2MASS photometry derived from the GNIRS spectrum described in $\S$\ref{sec:spec} yields an absolute $K_s$ magnitude of $M_{K_s} = 10.79$ and $J - K_s = 1.38$. On the $M_K$$_s$ versus $J - K_s$ CMD shown in Figure \ref{fig:CMD}, HIP 17453 B is photometrically consistent with an early L–dwarf \citep[e.g.][]{dup12}. 

\begin{figure}
    \gridline{\fig{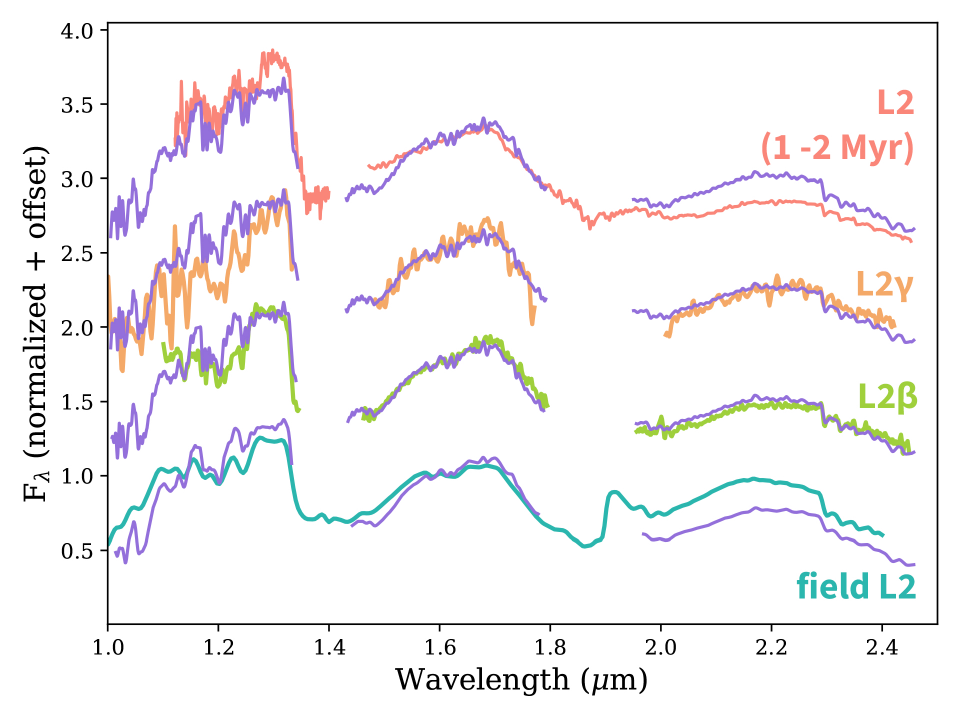}{0.75\textwidth}{}}
    \vspace{-1cm}
\caption{{GNIRS spectrum of HIP 17453 B (purple curves), plotted against an age sequence of a $\sim$1–2 Myr L2 brown dwarf \citep{patience12}, a $\sim$10 Myr L2$\gamma$ brown dwarf (orange curve) and a $\sim$100 Myr L2$\beta$ brown dwarf (green curve) \citep{gagne15}, and a $\gtrsim$1 Gyr field-age brown dwarf (salmon curve) \citep{cruz18}. All spectra are normalized to the peak H-band flux between 1.67 $\mu$m  $\leq \lambda \leq$ 1.68 $\mu$m . The spectrum of HIP 17453 B is most consistent with the L2$\beta$ spectrum.}}
    \label{fig:L2comp}
    \end{figure}

{Figure \ref{fig:L2comp} compares the GNIRS spectrum of HIP 17453 B against the spectra of L2 brown dwarfs with ages of $\sim$1–2 Myr, $\sim$10 Myr (L2$\gamma$), $\sim$100 Myr (L2$\beta$), and $\gtrsim$1 Gyr (field L2) from \citet{patience12}, \citet{gagne15} and \citet{cruz18}. The spectra of the youngest and oldest brown dwarfs do not match the spectrum of HIP 17453 B, showing a clear mismatch across $K$-band and in the $H$-band continuum slope. While the spectrum of HIP 17453 B shows good agreement with the $\sim$10 Myr brown dwarf, the $H$-band continuum slope is more triangular in the L2$\gamma$ object than in HIP 17453 B, which indicates this comparison object is too young or has a lower surface gravity than HIP 17453 B, as we expected with our calculated age of the primary. The spectrum of HIP 17453 B most closely resembles that of the $\sim$100 Myr brown dwarf (L2$\beta$). In particular, the $H$-band continuum matches the triangular shape and "shoulder" at 1.57 $\mu$m indicative of an intermediate-gravity object \citep{allers13}. As shown in Table \ref{tab:indices} and described in Section \ref{sec:spty}, the surface-gravity-sensitive features such as FeH and K I (as well as the $H$-cont index) resulted in a classification of intermediate gravity for the spectrum of HIP 17453 B, which is equivalent to the $\beta$ classification defined by \citet{cruz09} \citep{kirk10,allers13}. These spectral features and classifications support the interpretation that HIP 17453 B is younger than the typical field-age brown dwarf, which in turn supports the intermediate mass and gravity calculated for the object, positioning it as a unique substellar object of intermediate age and mass that bridges the gap between the lowest-mass stellar companions and highest-mass directly imaged planetary companions.}

\subsection{Mass Ratio}

\begin{figure}
    \gridline{
     \fig{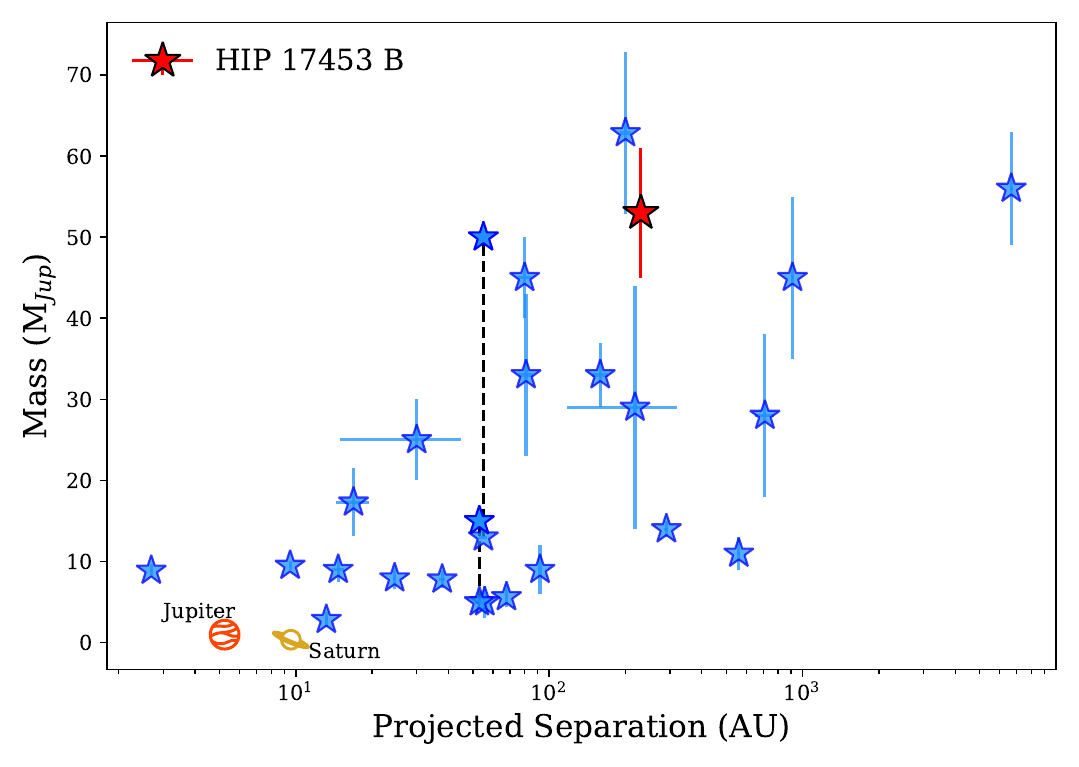}{0.65\textwidth}{}}
     \vspace{-10mm}
      \gridline{\fig{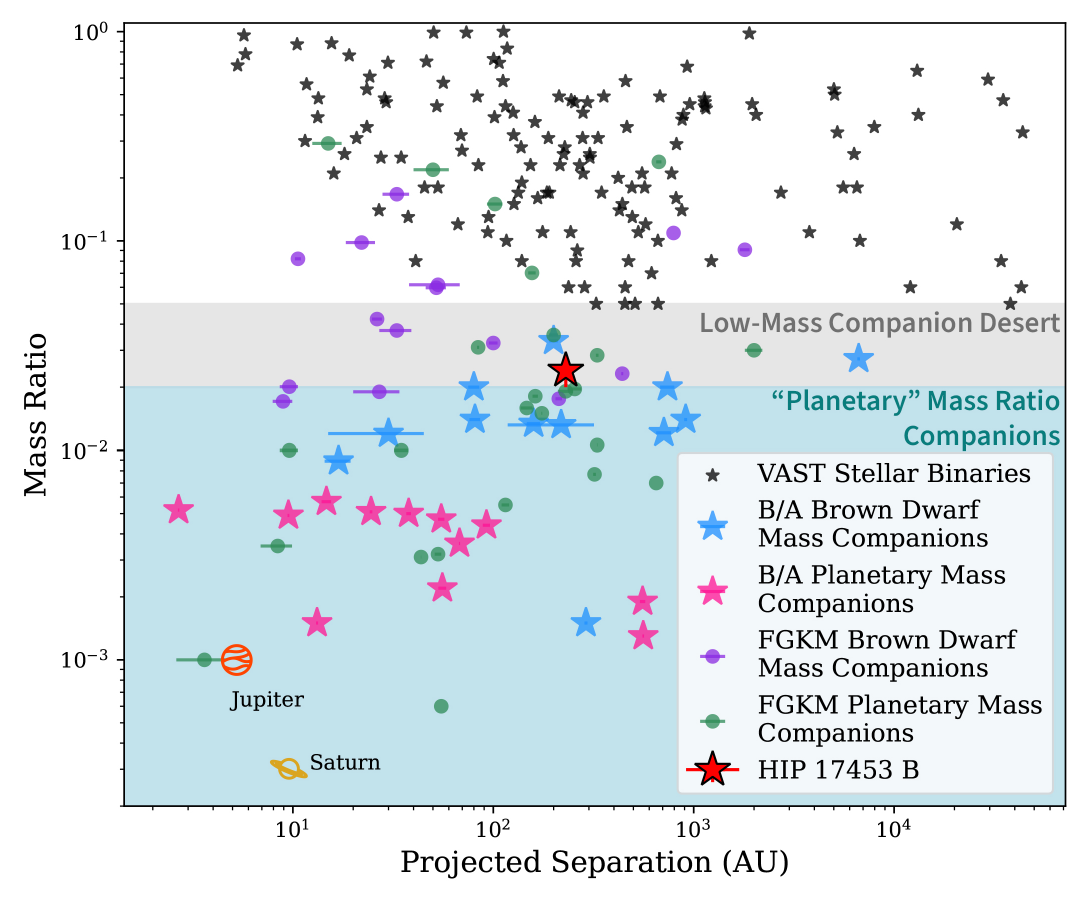}{0.65\textwidth}{}}
        \vspace{-1cm}
    \caption{(\textit{top}) Mass in M$_{Jup}$ as a function of separation in AU for the currently known directly imaged companions to B and A stars. HIP 17453 B is shown in red, and joins a small number of other directly imaged, wide separation brown dwarf companions {to early-type stars}. The uncertainty for the masses of $\kappa$ And b and HD 100546 b are shown with the dashed lines. (\textit{bottom}) Mass ratio as a function of separation for directly imaged brown dwarf and planetary mass companions to early-type (B/A) and FGKM stars, given by \citet{best2024}, as well as stellar binaries {with A-star hosts discovered by the Volume-limited A-STar (VAST) survey} from \citet{VAST3} (black stars). Companions around B and A stars are shown with the star symbol, while companions around FGKM stars are shown with the points. The gray shaded region denotes the low-mass companion desert to early-type stars, and the blue region denotes the regime of ``planetary'' mass ratio companions, both defined by \citet{duchene23}. HIP 17453 B (red star) is among the most widely separated, lowest mass ratio brown dwarf companions imaged to date, and straddles the boundary between the low-mass companion desert and where planetary mass ratio companions are found. 
    }
    \label{fig:MR}
    \end{figure}

HIP 17453 B joins only a handful of directly imaged known brown dwarf companions at wide separations around early-type stars, shown in the top panel of Figure \ref{fig:MR}. These objects, now including HIP 17453 B, occupy a sparsely populated segment of companion demographics for which it is critical to gain more data to fully probe the formation pathways for low-mass companions around higher-mass stars.
Based on the mass for HIP 17453 A, calculated as described in \S \ref{sec:host}, and the mass inferred for HIP 17453 B as described in \S\ref{sec:mass+teff}, the HIP 17453 system has a mass ratio of $q = 0.024 \pm 0.004$. This value places HIP 17453 B on the cusp of the ``low mass companion desert'' to intermediate-mass stars ($0.02 \leq q \leq 0.05$ for stars of mass $\sim$1.75-4.5 $M_{\odot}$) and ``planetary" mass ratio companions ($q \leq 0.02$) as defined by \citet{duchene23} and shown in the bottom panel of Figure \ref{fig:MR}. Companions around intermediate-mass stars were found to be underrepresented in the $0.02 \leq q \leq 0.05$ mass ratio range. This was interpreted as a "scaled-up" analogue to the brown dwarf desert around solar-type stars, with the deficit of these companions attributed to the inefficiencies of core accretion and disk fragmentation at producing extreme-mass-ratio binaries in the intermediate-stellar-mass regime. The placement of HIP 17453 B on the boundary between the two mass ratio populations provides a rare empirical data point that will probe the transition between brown-dwarf-mass and planetary-mass companions around intermediate-mass stars. Given the relative deficit of brown-dwarf-mass companions in this mass ratio range, especially compared to the stellar companions with larger mass ratios and planetary companions with lower mass ratios, suggests that this analogue brown dwarf desert may be less sharply defined at its lower boundary than previously assumed, and instead represents a steep decline rather than a strict exclusion zone \citep{duchene23}. The discovery of HIP 17453 B supports the growing evidence that companions at or just above the planetary–brown-dwarf-mass boundary are very rare around intermediate-mass stars, but not nonexistent. This object provides an important calibration point for companion demographics in this sparsely-populated mass ratio regime where the formation pathways for substellar companions remain uncertain. The location of HIP 17453 B at the boundary between the low-mass companion desert and the planetary mass ratio domain makes it a valuable test case for distinguishing between competing formation scenarios such as turbulent core fragmentation or gravitational instability in circumstellar disks. Continued discovery and characterization of systems like HIP 17453 B will be essential for refining our understanding of the statistical properties of substellar companions, particularly across the poorly constrained transition from planetary to brown dwarf masses.

\subsection{Companion Demographics}

\subsubsection{Preliminary Results from the C-BASS Survey}

The C-BASS Survey has observed 211 stars, and has thus far detected one brown dwarf companion to a single star. Analysis of candidate companions is still ongoing, and so it is possible that more brown dwarf companions near the edge of the contrast curve will be detected from this analysis. In a future paper we will examine the companion demographics more robustly after considering the completeness to planets for each target star \citep[e.g.][]{GPIES,vigan21}. For now, we consider the implications of a $\sim$0.5-2\% detection rate of wide-separation brown dwarfs around intermediate mass stars. This detection rate can be estimated by the following equation:

\begin{equation}
  N_{det} \sim \frac{N_{BD}}{N_{Stars}}
\end{equation}
where $N_{BD}$ is the number of brown dwarfs detected, and $N_{Stars}$ is the total number of host stars observed.

The Gemini Near-Infrared Coronagraphic Imager (NICI) Planet-Finding Campaign \citep{liu10} observed 70 B and A stars and detected four brown dwarfs around three stars: HR 7329 B ($\eta$ Tel B), HD 1160 B, and HIP 79797 Ba/Bb (HR 6037 Ba/Bb) \citep{nici2}. These objects have masses of $47^{+5}_{-6}$M$_{{Jup}}$ \citep{lazzoni20}, 16-81 M$_{{Jup}}$ \citep{sutlieff24}, $58^{+21}_{-20}$M$_{{Jup}}$, and $55^{+20}_{-19}$M$_{{Jup}}$ \citep{nici2}, respectively. All three host stars are likely members of moving groups: HR 7329 is in the $\beta$ Pic moving group and so has a relatively certain age, HD 1160 is a likely member of the Pisces-Eridanus stream \citep{curtis19}, and HIP 79797 is a member of Argus \citep{zuckerman11}. Using a younger age for HIP 79797 (40-50 Myr) given its Argus membership, the masses of both components shift downward to $\sim$30 M$_{{Jup}}$. The projected separations for these four companions are $\sim$4.2$^{\prime\prime}$ ($\sim$200 AU) \citep{lowrance00}, $\sim$0.8$^{\prime\prime}$ ($\sim$80 AU) \citep{nielsen12}, and $\sim$6.7$^{\prime\prime}$ ($\sim$370 AU) \citep{huelamo10}, respectively. HIP 79797 Ba/Bb in turn is a binary brown dwarf with a separation between the components of 60 mas ($\sim$3 AU) \citep{nici2}.

In general, brown dwarfs similar to these should have been detectable in the C-BASS Keck/NIRC2 observations. Even 5-minute snapshots reach most brown dwarf masses beyond $\sim$1$^{\prime\prime}$, and the field of view is complete out to $\sim$4$^{\prime\prime}$. Thus is it possible some analogs to HD 1160 B might have been too close to distant stars to be detectable, or HIP 79797 Ba/Bb analogs might have been too far from nearby stars to be within the field of view. Nevertheless, the difference in the implied occurrence rate between NICI and C-BASS is dramatic: 4/70 = 6\% and 1/211 = 0.5\%, respectively. The deeper NICI observations (which were more sensitive to closer-in companions) and the wider field of view ($\sim$10$^{\prime\prime}$ in radius instead of $\sim$5$^{\prime\prime}$) may explain some or all of the discrepancy. A full demographics analysis of the C-BASS survey (as well as final vetting of all candidate companions) will be necessary to truly compare the two surveys.

\subsubsection{Implications for the Occurrence Rate of Companions to B and A Stars}

As a preliminary step we can compare existing demographics of stellar, brown dwarf, and giant planet companions from the literature. We begin with lower-mass stellar binaries, using The Volume-limited A-STar (VAST) survey \citet{VAST3}. VAST detected 39 companions with masses between  75 M$_{{Jup}}$ – 0.5 M$_\odot$, with projected separations between 100--1000 AU, out of 363 stars, with close to 100\% completeness over this full range. This implies an occurrence rate of approximately 10\%. VAST also found 13 companions over a similar mass range between 10–100 AU, but with completeness varying from 100\% (at larger masses and larger separations) to 0\% (at the smallest masses and smallest separations). We estimate the average completeness across this range to be 50\%, which then implies an occurrence rate of $\sim$7\%.

For the occurrence rate of brown dwarfs between 10–100 AU we use the GPIES results from \citet{GPIES}. GPIES detected 0 brown dwarfs around stars above 1.5 M$_\odot$ (corresponding to a main sequence spectral type of F1), and had an average completeness of 80 stars from 10-100 AU, 13-75 M$_{{Jup}}$. The implied upper limit on occurrence rate from this non detection is $\lesssim$1.2\%. We note this analysis was performed over semi-major axis rather than separation, but we equate the two to make these estimates. This region of parameter space is clearly not empty, with two objects within this mass limit having been recently detected by a SCExAO survey of \texttt{Hipparcos}/\texttt{Gaia} accelerating stars. HIP 99770 B is 13.9–16.1 M$_{{Jup}}$ at 17 AU \citep{currie23} and HIP 39017 B is $\sim$30 M$_{{Jup}}$ at 24 AU \citep{tobin24}. Deriving occurrence rates from astrometrically selected samples is difficult, but taken together these surveys suggest a low but non-zero occurrence rate of intermediate separation brown dwarfs around intermediate mass stars.

Over the same mass range and for semi-major axis between 100–1000 AU, the NICI Campaign detected 3 brown dwarfs (not including HD 1160 B at 80 AU) with an average completeness of 44 stars \citep{nici2}. As a result, the occurrence rate this suggests for wider-separation brown dwarfs is $\sim$7\%. The SHINE survey \citep{vigan21} also probed this region, finding 3 brown dwarfs around 53 B and A stars. Completeness is likely between 50\%–100\% based on the reported depth of search plot, consistent with a $\sim$6–12\% occurrence rate.

Finally, we consider giant planets over this same range of semi-major axis. The reported occurrence rate from GPIES of 5–13 M$_{{Jup}}$ planets from 10–100 AU is $8.9^{+4.0}_{-4.6}$\%. The SHINE campaign detected five planets in this range from 53 B and A stars, and for a completeness between 50\% to 100\% that corresponds to an occurrence rate of $\sim$9\%–18\%. For wider separation planets, between 100–1000 AU a single brown dwarf was detected. For a similar assumption of completeness between 50\%–100\%, that implies an occurrence rate of $\sim$2\%–4\%.

\begin{figure}
    \gridline{\fig{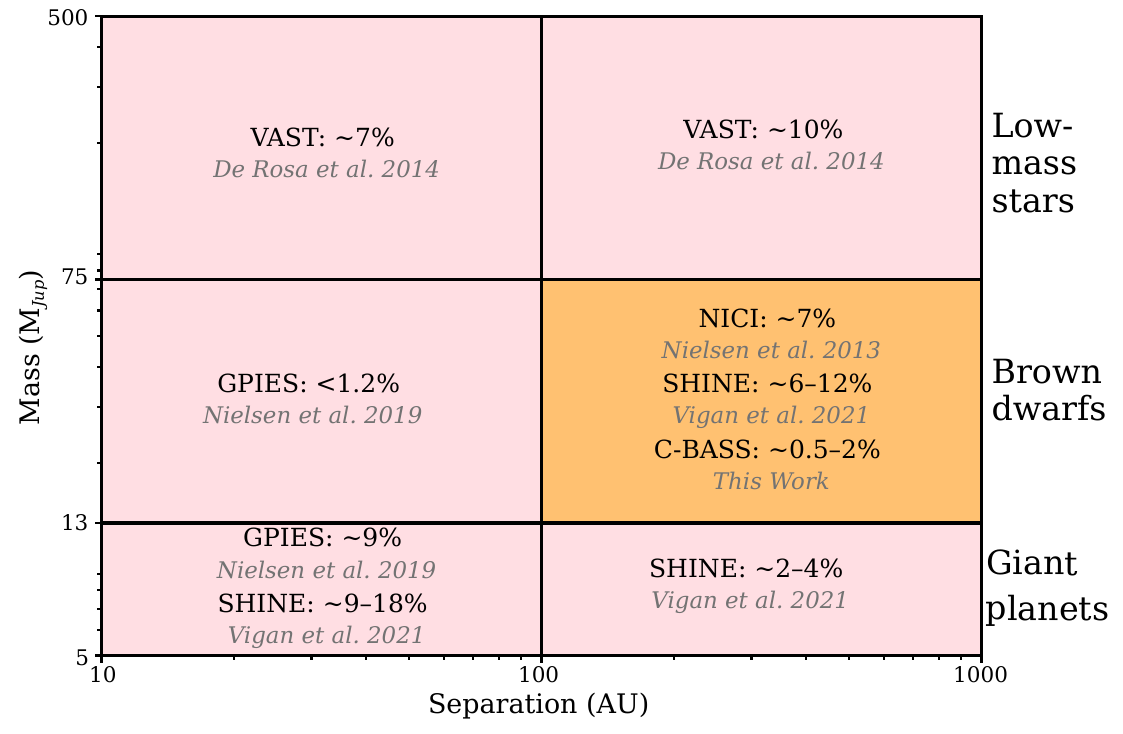}{0.75\textwidth}{}}
        \vspace{-1cm}
    \caption{Approximate occurrence rates in log-log space for planets, brown dwarfs, and M-dwarfs at different separations around B and A star hosts. Between 10-100 AU these results are consistent with a much lower number of brown dwarf companions compared to M-star binary companions and giant planets, perhaps pointing to different formation mechanisms for these populations. At wider separations occurrence rates are more similar for brown dwarfs and M-star companions, then dropping slightly for planets. There is a discrepancy between the low C-BASS occurrence rate and the higher NICI and SHINE occurrence rates. This may be due to a change in occurrence rate of wide-separation brown dwarf binaries with stellar age, though a more robust demographics analysis is required to properly map out these trends.
    }
    \label{fig:notatongue}
    \end{figure}

While these are estimated values and not a rigorous demographics analysis, we display these approximate occurrence rates schematically in Figure~\ref{fig:notatongue}. We also note the mass ranges are not equal: while the low-mass stars and brown dwarfs are roughly equally spaced in log, the range in log-mass covered by the planet bin is about a factor of two smaller. Thus the planet occurrence rates represent a lower limit for a similar range of masses (which, to match, would require a mass range of 2–13 M$_{\textrm{Jup}}$).

For closer separations (10–100 AU) there is a suggestion of a brown dwarf desert around B and A stars: $\sim$10\% occurrence rates for both low-mass stellar companions and giant planets, and an order of magnitude smaller occurrence rate for brown dwarfs. A separate trend emerges at wider separations, with occurrence rate staying roughly constant (maybe decreasing slightly) from stellar companions to brown dwarfs, then continuing to decrease when moving to giant planets. The precise shape of this trend will again require a more robust demographic analysis, as well as turn on the final brown dwarf yield of the C-BASS campaign.

These preliminary results, however, are in line with demographics analyses that suggest intermediate-separation ($\sim$10–100 AU) brown dwarfs and giant planets are not the same populations \citep{GPIES,bowler20}. Indeed, from this simplified analysis, the giant planet population and brown dwarf companion populations to B and A stars appear to be inverted. Brown dwarf companions appear to be an order of magnitude more common at 100–1000 AU compared to 10–100 AU. Meanwhile giant planets appear several times more numerous at 10–100 AU compared to 100–1000 AU.

One possible explanation for the divergent demographics between brown dwarfs and giant planets around B and A stars from 10–1000 AU is formation. In particular, that binary stars and planets are two separate formation processes, with brown dwarfs representing some mix of low-mass stars and high-mass planets. Brown dwarfs may even represent a third formation mechanism (e.g. gravitational instability, \citealt{GPIES}). Alternatively, this may be an effect of dynamical evolution, with the observed population representing snapshots in time as these companions migrate.

Indeed, one potential contribution to the lower brown dwarf occurrence rate from C-BASS compared to NICI and SHINE is stellar age. C-BASS stars tend to be older than the NICI and SHINE samples, which contain more moving group B and A stars. All of the NICI and SHINE brown dwarf detections are members (or proposed members) of moving groups and associations: HR 7329 is a member of the $\beta$ Pic moving group \citep{lazzoni20}, HIP 64892 is a member of Lower Centaurus Crux \citep{cheetham18},  HIP 78530 is a member of Upper Scorpius \citep{lafreniere11}, HIP 79797 is a member of Argus \citep{zuckerman11}, and HD 1160 is a possible member of the Pisces-Eridanus stream \citep{curtis19}. Wide-separation binaries with a brown dwarf component have lower binding energies than closer or more massive binaries. Thus, one resolution to the occurrence rate discrepancy is that these wide ($\gtrsim$100 AU) brown dwarf binaries become unbound after $\sim$100 Myr. Another explanation is that since moving group stars (like the hosts of the brown dwarfs detected by NICI and SHINE) form in low density star-forming environments, these stars are more likely to maintain wide-separation companions than the stars formed in denser environments, as with the stars in the C-BASS sample, where companion formation and retention at wide separations could be suppressed due to nearby stellar neighbors. In any case, a more complete demographics analysis is needed to determine how robust these occurrence rate trends truly are.

\section{Summary and Conclusions} \label{sec:summ}

We have reported the discovery of a brown dwarf companion to the chemically peculiar field A0-star HIP 17453 A at a projected separation of {230}AU. With an age estimate of 280 ± 125 Myr for {HIP 17453 A}, we determine a mass for HIP 17453 B of  $53^{+10}_{-8}$ M$_{{Jup}}$ using the Sonora Diamondback \citep{sonoradiamondback} substellar evolution model grids. Analysis of the spectrum yielded a spectral type of L2 $\pm$ 1, with $T_{eff}$ $\simeq$ 1900 K and $log(g)$ between 5.0 and 5.5 based on comparisons to the BT-Settl and Sonora Diamondback atmosphere models. The HIP 17453 system has an extreme mass ratio of $q = 0.024 \pm 0.004$, placing it on the border between the low-mass companion desert for intermediate mass stars and planetary-mass-ratio companions as defined by \citet{duchene23}. With {a well-constrained} age {of 280 ± 125 Myr, HIP 17453 B serves as a critical benchmark for brown dwarf evolutionary models, bridging the gap between young moving groups/star-forming regions and  and the mature field star population.} 

With the confirmation of HIP 17453 B as a brown dwarf companion to an A-star, future studies of this object will provide important insights into the mechanisms governing substellar formation and evolution, particularly in comparison to {directly-imaged} planetary systems. HIP 17453 B is a valuable system for follow-up measurements with high-resolution spectroscopy and continued orbital monitoring. AO-fed high resolution spectrographs such as the Keck Planet Imager and Characterizer (KPIC, \citealt{kpic2,kpic3}) can be used to measure the rotational velocity \textit{v} sin (\textit{i}) for the substellar companion (e.g. \citealt{kpic}). In general, \textit{v} sin (\textit{i}) can offer a diagnostic of the evolutionary history of an object, and is influenced by factors such as contraction and mass loss rate. For solar-type stars, \citet{sku72} found that the rotational velocity rates are strongly dependent on the age of the star; their rotational velocities decrease (or spin down) due to loss of angular momentum and mass via stellar winds following magnetic field lines as the star ages. On the other hand, planetary-mass companions show uniformly low rotational velocities that are not correlated with age; these objects instead decrease in rotational velocity due to interactions with the protoplanetary disk in which they are forming \citep{bryan18}. By comparison,  the evolution of rotational velocity rates for substellar objects over time is less well-calibrated compared to their stellar and planetary counterparts \citep{jo03,zo06,bryan18}, indicating a need for measurements at intermediate masses and ages. Rotational evolution is especially poorly sampled at the age of the HIP 17453 system. Measuring the $\textit{v} sin (\textit{i})$ of HIP 17453 B will provide a benchmark measurement for the spin evolution for objects with intermediate ages between star-forming regions and field stars. 

Higher signal-to-noise spectra, preferably with the wavelength range extended into the mid-infrared, would also provide measurements of elemental abundances (e.g., C/O ratio and metallicity) in the atmosphere of the brown dwarf, which are predicted to depend on the formation history and composition of the object (e.g. \citealt{oberg2011}). For example, recent analysis of the brown dwarf companion HD 33632 Ab revealed elemental abundances consistent with those of its host star \citep{kpic}, contrasting with many hot Jupiters that frequently exhibit enhanced metal enrichment from core accretion processes \citep{morb23}. Recent JWST/MIRI data of Gliese 229 Bab reveal the C/O and metallicity are consistent with stellar values, reinforcing the case for formation via gravitational collapse rather than significant alteration through disk processing \citep{xuan24}. With similar measurements for HIP 17453 B, we can evaluate the likely formation history of the object and perform a detailed analysis of the brown dwarf atmosphere. 

Through orbital monitoring, the spin-orbital alignment (or spin-orbit) of binary components trace the formation and dynamical history of the system. A-star primaries with close-orbit ($<$5AU) companions show distinct patterns in obliquity for star-star \citep{smith24} and star-planet \citep{winn15} systems. The circularized stellar pairs typically exhibit spin-orbit alignment at low mass ratios and eccentricities, while the star-planet systems reveal a wide range of angles between the stellar spin axis and orbital plane axis. For wider pairs, longer term orbital motion monitoring is required to determine the inclination of the orbital plane, however measurements of the spin-orbit angles for populations of both binary stars \citep{hale94} and star-substellar systems \citep{bowl23} have been measured. Both types of systems were aligned for separations within tens of AU, and showed a wide range of angles between the spin axis and orbital axis for wider pairs. The inferred mass of HIP 17453 B is not well sampled in the star-substellar pairs, and, with continued monitoring, could add to the population statistics. The separation of 2.9$^{\prime\prime}$ means HIP 17453 would be an ideal system to follow-up with high resolution spectroscopy and photometry. The spin-orbit of the HIP 17453 system could be measured using spectro-interferometry, as was demonstrated with the $\beta$ Pictoris system \citep{kraus2020}.
With its distinctive combination of intermediate mass and age and its accessibility for follow-up, HIP 17453 B joins a small set of benchmark brown dwarfs at the boundary between stars and planets amenable for comparative atmospheric and evolutionary studies. 

\begin{acknowledgments}
\section*{Acknowledgments}
This research has made use of the Keck Observatory Archive (KOA), which is operated by the W. M. Keck Observatory and the NASA Exoplanet Science Institute (NExScI), under contract with the National Aeronautics and Space Administration.

Some of the data presented herein were obtained at Keck Observatory, which is a private 501(c)3 non-profit organization operated as a scientific partnership among the California Institute of Technology, the University of California, and the National Aeronautics and Space Administration. The Observatory was made possible by the generous financial support of the W. M. Keck Foundation. 

This material is based upon work supported by NASA under Grant No. 80NSSC22K0485 issued through the Science Mission Directorate Astrophysics Division Astrophysics Data Analysis Program. AW also acknowledges support from NASA Grant No. 80NSSC23K1356 (PI: Steve Desch) issued through the Science Mission Directorate Planetary Science Division. EN acknowledges support from NSF Grant No. AST07-09484 through the National Science Foundation. 

This work was supported by a NASA Keck PI Data Award, administered by the NExScI. Data presented herein were obtained at the W. M. Keck Observatory from telescope time allocated to the National Aeronautics and Space Administration through the agency's scientific partnership with the California Institute of Technology and the University of California. The Observatory was made possible by the generous financial support of the W. M. Keck Foundation.

This work was enabled by observations made from the W. M. Keck Observatory telescope and Gemini North telescope, located within the Maunakea Science Reserve and adjacent to the summit of Maunakea. We are grateful for the privilege of observing the Universe from a place that is unique in both its astronomical quality and its cultural significance. The authors wish to recognize and acknowledge the very significant cultural role and reverence that the summit of Maunakea has always had within the Native Hawaiian community. We are most fortunate to have the opportunity to conduct observations from this mountain. 

This study uses data processed with the Gemini IRAF package, based on observations obtained at the international Gemini Observatory, a program of NSF NOIRLab, which is managed by the Association of Universities for Research in Astronomy (AURA) under a cooperative agreement with the U.S. National Science Foundation on behalf of the Gemini Observatory partnership: the U.S. National Science Foundation (United States), National Research Council (Canada), Agencia Nacional de Investigación y Desarrollo (Chile), Ministerio de Ciencia, Tecnología e Innovación (Argentina), Ministério da Ciência, Tecnologia, Inovações e Comunicações (Brazil), and Korea Astronomy and Space Science Institute (Republic of Korea).

NOIRLab IRAF is distributed by the Community Science and Data Center at NSF NOIRLab, which is managed by the Association of Universities for Research in Astronomy (AURA) under a cooperative agreement with the U.S. National Science Foundation.

We would like to thank the students of SES 294 for their contributions to the observations used in this work: M. Antares, J. Ball, E. Catogni, K. Chauhan, R. Cochran-White, J. Colborn, S. Corley, K. Creel, K. El-Wattar, B. Hager, A. Humphrey, N. Jeantilus, A. Johnson, S. Marquez-Perez, C. Merchant, K. Munoz, F. Noguer, K. Norton, E. Poniewaz, K. Reagan, T. Skrmetti, D. Wood, B. Zewe. 

We thank the reviewer for their constructive comments and suggestions which greatly improved the quality of this manuscript.

 \end{acknowledgments}

\vspace{5mm}
\facilities{Keck II/NIRC2, Gemini-North/GNIRS, Keck Observatory Archive (KOA), Gemini Observatory Archive}

\software{\texttt{pyKLIP} \citep{pyklip}, \texttt{astropy} \citep{astropy2,astropy1}, \texttt{species} \citep{species}, \texttt{orbitize!} \citep{blunt2017,blunt2019}, \texttt{zeus} \citep{zeus}}
    
\bibliography{ref}
\bibliographystyle{aasjournal}



\end{document}